\documentclass[longauth]{aa}  
\usepackage{graphicx}
\usepackage{amsmath}
\usepackage{wasysym}
\usepackage{mathtools} 
\usepackage{booktabs} 
 \usepackage{rotating,multirow} 
\usepackage[UKenglish]{isodate}
\cleanlookdateon 
\usepackage{gensymb}
\usepackage{graphicx}
\usepackage{upquote}

\usepackage{natbib,twoopt}
\usepackage[breaklinks=true]{hyperref} 
\bibpunct{(}{)}{;}{a}{}{,} 
\makeatletter

\usepackage{color}
\usepackage{soul}

\graphicspath{{./figures/}}

\def\teff{$T_{\rm eff}$}

\def\bp{$G_{\rm BP}$}
\def\rp{$G_{\rm RP}$}
\def\bmr{$(G_{\rm BP}-G_{\rm RP})$}
\def\mg{$M_G$}

\def\gaia{\textit{Gaia}}

\begin{document} 

\title{\gaia\ Data Release 2}
\titlerunning{Gaia DR2: Rotational modulation in late-type dwarfs}
\subtitle{Rotational modulation in late-type dwarfs}

\author{
A.~C. Lanzafame\inst{1,2} \and 
E. Distefano\inst{2}      \and
S. Messina\inst{2}          \and
I. Pagano\inst{2}     \and
A.~F. Lanza\inst{2}        \and
L. Eyer\inst{3,4}         \and
L.~P. Guy\inst{4}        \and
L. Rimoldini\inst{4}       \and
I. Lecoeur-Taibi\inst{4}        \and     
B. Holl\inst{3,4}        \and
M. Audard\inst{3,4}      \and
G.~J. de~Fombelle\inst{4}        \and
K. Nienartowicz \inst{4}          \and
O. Marchal \inst{4} \and
N. Mowlavi\inst{3,4}        
}

\authorrunning{A.~C. Lanzafame et al.}

\offprints{A.~C. Lanzafame \\ \email{a.lanzafame@unict.it}}

\institute{
Universit\`a di Catania, Dipartimento di Fisica e Astronomia, Sezione Astrofisica, Via S. Sofia 78, I-95123 Catania, Italy \\
 \email{a.lanzafame@unict.it}
\and
INAF-Osservatorio Astrofisico di Catania, Via S. Sofia 78, I-95123 Catania, Italy
\and
Department of Astronomy, University of Geneva, Chemin des Maillettes 51, 1290 Versoix, Switzerland 
\and
Department of Astronomy, University of Geneva, Chemin d'Ecogia 16, 1290 Versoix, Switzerland 
}

\date{Received 2 May 2018 ; Accepted 23 May 2018}

\abstract{
Amongst the $\approx 5\times 10^5$ sources identified as variable stars in \gaia\ Data Release 2 (DR2), 26\% are rotational modulation variable candidates of the BY\,Dra class.
\gaia\ DR2 provides their multi-band ($G$, \bp\ , and \rp ) photometric time series collected by the European Space Agency spacecraft \gaia\ during the first 22 months of operations as well as the essential parameters related to their flux modulation induced by surface inhomogeneities and rotation.
}{
We developed methods to identify the BY\,Dra variable candidates and to infer their variability parameters. 
}{
BY\,Dra candidates were pre-selected from their position in the
Hertzsprung-Russel diagram, built from \gaia\ parallaxes, $G$ magnitudes, and \bmr\ colours.
Since the time evolution of the stellar active region can disrupt the coherence of the signal, segments not much longer than their expected evolution timescale were extracted from the entire photometric time series, and period search algorithms were applied to each segment. 
For the \gaia\ DR2, we selected sources with similar periods in at least two segments as candidate BY\,Dra variables.
Results were further filtered considering the time-series phase coverage and the expected approximate light-curve shape.
}{
\gaia\ DR2 includes rotational periods and modulation amplitudes of 147\,535 BY\,Dra candidates.
The data unveil the existence of two populations with distinctive period and amplitude distributions.
The sample covers 38\% of the whole sky when divided into bins (HEALPix) of $\approx$0.84 square degrees, and we estimate that this represents 0.7 -- 5 \% of all BY\,Dra stars potentially detectable with \gaia.
}{
The preliminary data contained in \gaia\ DR2 illustrate the vast and unique information that the mission is going to provide on stellar rotation and magnetic activity.
This information, complemented by the exquisite \gaia\ parallaxes, proper motions, and astrophysical parameters, is opening new and unique perspectives for our understanding of the evolution of stellar angular momentum and dynamo action.
}


\maketitle


\section{Introduction}
\label{sec:Introduction}

\begin{figure}[t]
\begin{center}
\includegraphics[width=0.44\textwidth]{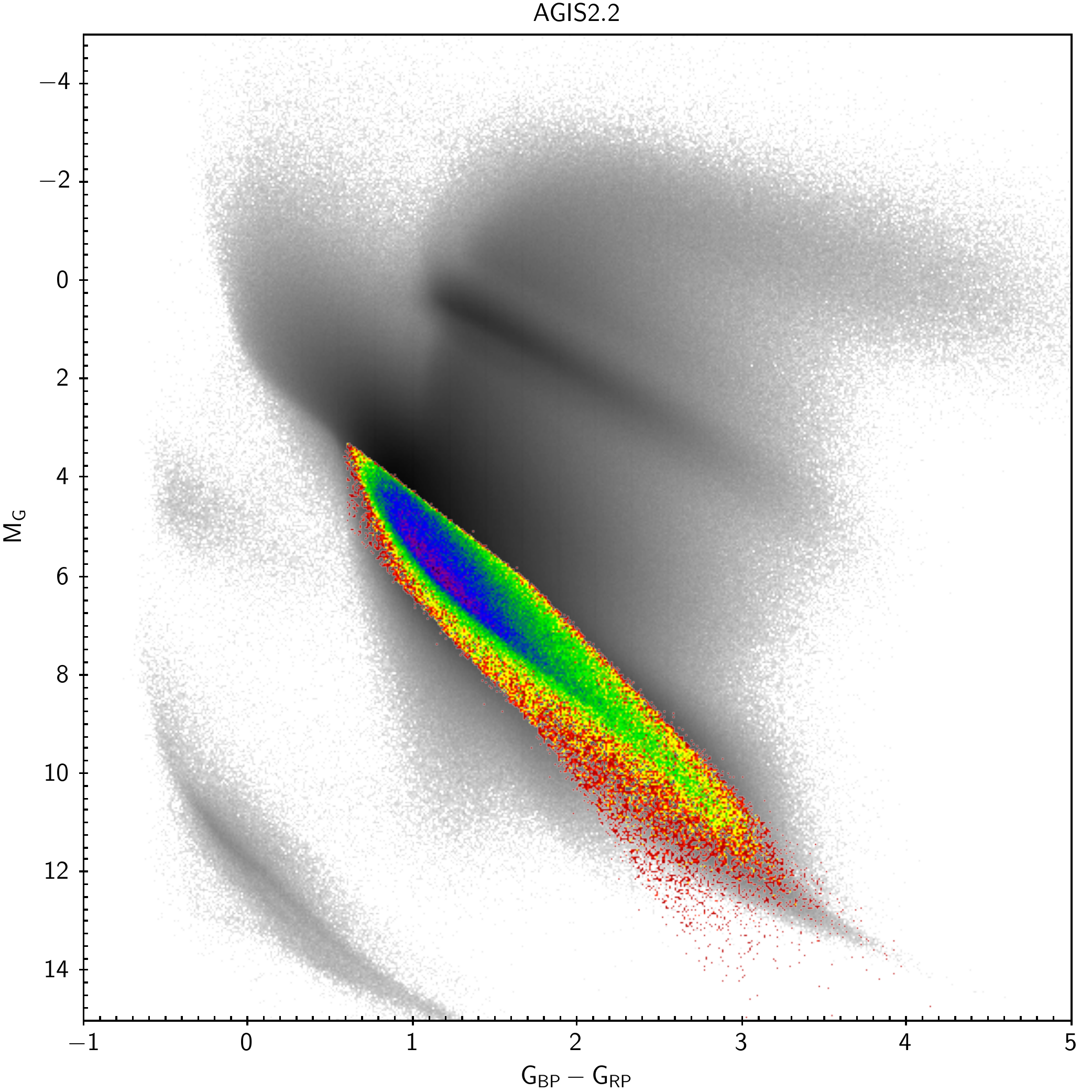}
\caption{Region of the HRD in which rotational modulation was found. 
The region is highlighted with colours (rainbow). The colour code indicates the relative density of data points (red for lower
density, and purple for higher density).
The HRD shown is built using the AGIS02.2 solution for the parallax \citep[left panel, see][for details]{2018arXiv180409366L} uncorrected for interstellar extinction.}
\label{fig:hr}
\end{center}
\end{figure}

\gaia\ all-sky scanning and multi-epoch observations offer unprecedented opportunities to detect and characterise stellar variability.
One of the variability phenomena that can be studied using \gaia\ data is the flux modulation induced by surface inhomogeneities (such as spots and faculae) combined with rotation. An analysis of this modulation allows us to infer the stellar rotation period.
In this paper we refer to dwarf, sub-giant, and T\,Tauri stars showing such flux modulation as BY\,Dra variables, after the archetype BY\,Draconis.

As for the Sun, the variability in late-type stars also includes more irregular effects that are due to the emission originating from the chromosphere/corona, and to sporadic outbursts that
in turn are due to reconnection of magnetic flux tubes, that
is, flares. 
Numerous classes of stars (single and multiple systems) are affected by this variability, including BY\,Dra, UV\,Cet, and RS\,CVn types.
As in classical papers on the subject, this classification is phenomenological: when flares are observed, the star is classified to be of UV\,Cet type; a BY\,Dra can be a single star or a star in a binary system; and when the star is in a close binary system, a condition that enhances magnetic activity, the system is classified to be of RS\,CVn type. 

Variability due to spots and faculae can reach amplitudes of the order of some tenths of magnitudes for the most active objects (e.g. T\,Tauri stars), but can be of the order of milli-magnitudes or lower for late-type dwarfs of solar age.
Inferring rotation periods and modulation amplitude in a large and heterogeneous sample of stars, both in clusters and in the field, for which other astrophysical parameters can also be inferred homogeneously, as in the \gaia\ case, opens new frontiers in our understanding of stellar angular momentum evolution.
Together with information from asteroseismology, the \gaia\ data and analysis provide previously unattainable insights that enable us to effectively study issues like the dependence of surface rotation on age, the existence of different dynamo regimes, and the role of multiplicity, planetary systems, and environment at large in the stellar angular momentum evolution.

As part of the activities carried out by the variability Coordination Unit (CU7) of the Gaia Data Processing and Analysis Consortium (DPAC), we have devised and implemented methods for the identification and analysis of rotational modulation in low-mass stars.
In this paper we present these methods and give an overview of the periods and modulation amplitudes obtained on the first 22-month \gaia\ baseline, which are available in \gaia\ DR2.
In Sect.\,\ref{sec:Method} we present the methods we developed, in Sect.\,\ref{sec:Results} we describe the results, and in Sect.\,\ref{sec:Conclusions} we draw our conclusions.

\section{Method}
\label{sec:Method}

The problem of inferring stellar rotation periods from \gaia\ photometric time series has been discussed in \cite{2012MNRAS.421.2774D}.
In brief, on the one hand, the \gaia\ sampling depends on the star's position on the sky and is irregular, in the sense that the time interval between two consecutive transits is not constant and rather sparse. There are, of course, regularities like those due to the consecutive passages of the star's image on the two satellite's FoVs and the precession of the satellite, but the time step is not constant. \gaia\ DR2 also contains a subset in which regions around the ecliptic poles were observed with a regular sampling.  
On the other hand, in analogy with the solar case, the distribution of inhomogeneities on the surface of a low-mass star surface changes in time as a result of the evolution of spots, faculae, and active region complexes, each with its own characteristic timescale, making the rotational modulation signal, in general, non-coherent over long timescales. 
Furthermore, the stochastic variability due to flaring is superimposed on the rotational modulation. 

\cite{2012MNRAS.421.2774D} adopted the detailed light-curve model of \cite{2006A&A...455..595L} for their simulation of \gaia\ observations of solar twins. 
Extrapolating these models to stars of different mass and/or magnetic activity is still very uncertain and not justified for the \gaia\ case because of the sparseness of its sampling.
We therefore do not attempt to fit a detailed model to the data, but rather search for periodicities in the light curves that can be produced by rotational modulation by assuming a sinusoidal model.
Since this can be applied to any (quasi-)periodic variability, it is necessary to constrain our classification using other information available in the \gaia\ data.

To achieve a reliable identification of stellar variability classes like that of BY\,Dra, we plan to combine the variability parameters with the \gaia\ inference of astrophysical parameters \citep{2013A&A...559A..74B}.
However, only preliminary astrophysical parameters based on integrated photometry are available in DR2 \citep{2018arXiv180409374A}, and therefore we here adopted a simplified approach (Sect.\,\ref{sec:InputData}).

Within the \gaia\ variability pipeline, rotational modulation is processed by two modules: the {\it \textup{special variability detection (SVD) solar-like}}, devised for a first selection of candidates, time-series segmentation, and outlier detection (possible flares), and the {\it \textup{specific object study} \textup{(SOS), rotational modulation}} devised for period search, modulation amplitude estimate and filtering of the final results.
The whole procedure is described in the following.

\subsection{Input data}
\label{sec:InputData}

We refer to \cite{2018arXiv180409373H} for a summary of the general variability processing, analysis, and the observation filtering and flagging applicable to the results published in \gaia\ DR2. 

The simulations presented in \cite{2012MNRAS.421.2774D} indicate that the analysis of rotational modulation in solar-like stars requires at least a dozen samplings in at least one interval not longer than $\sim$\,150\,d (see also Sect.\,\ref{sec:Segmentation}).
The rotational modulation processing, therefore, follows the geq20 path described in  \cite{2018arXiv180409373H}, that is, the path starting from the 826 million sources with $\ge$ 20 field-of-view (FoV) transits. 

From the initial geq20 catalogue, an initial list of BY\,Dra candidates was compiled by selecting sources around the observed main sequence in the (\mg, \bmr) diagram.
The approximate absolute magnitude \mg\ was estimated from parallax and $G$, ignoring reddening, to build the diagrams shown in Fig.\,\ref{fig:hr}.
Only sources with a relative parallax uncertainty better than 20\% were considered.
The selected region (coloured region in Fig.\,\ref{fig:hr}) broadly embraces the main sequence.
The leftmost boundary, \bmr=0.6, limits the selection to stars of spectral type approximately later than F5. 

The initial list of BY\,Dra candidates was then further limited to sources whose time series could be divided into at least two segments with a number of transits $N_G \ge 12$.
The segmentation algorithm is described in Sect.\,\ref{sec:Segmentation}.

\subsection{Time-series segmentation}
\label{sec:Segmentation}

\cite{2004A&A...425..707L} analysed the total solar irradiance (TSI) time series collected by the VIRGO experiment on the SoHo satellite and found that period-search algorithms can reproduce the correct solar rotation period only if the effects due to the evolution of the active region complexes are limited by dividing the long-term time series into sub-series of $\sim$150\,d. 
The active region evolution timescale in young stars, which have
a higher magnetic activity than the Sun, can be significantly shorter \citep[$\sim$60--90\,d,][]{2003A&A...410..671M}.

\begin{figure}
\begin{center}
\includegraphics[width=0.44\textwidth]{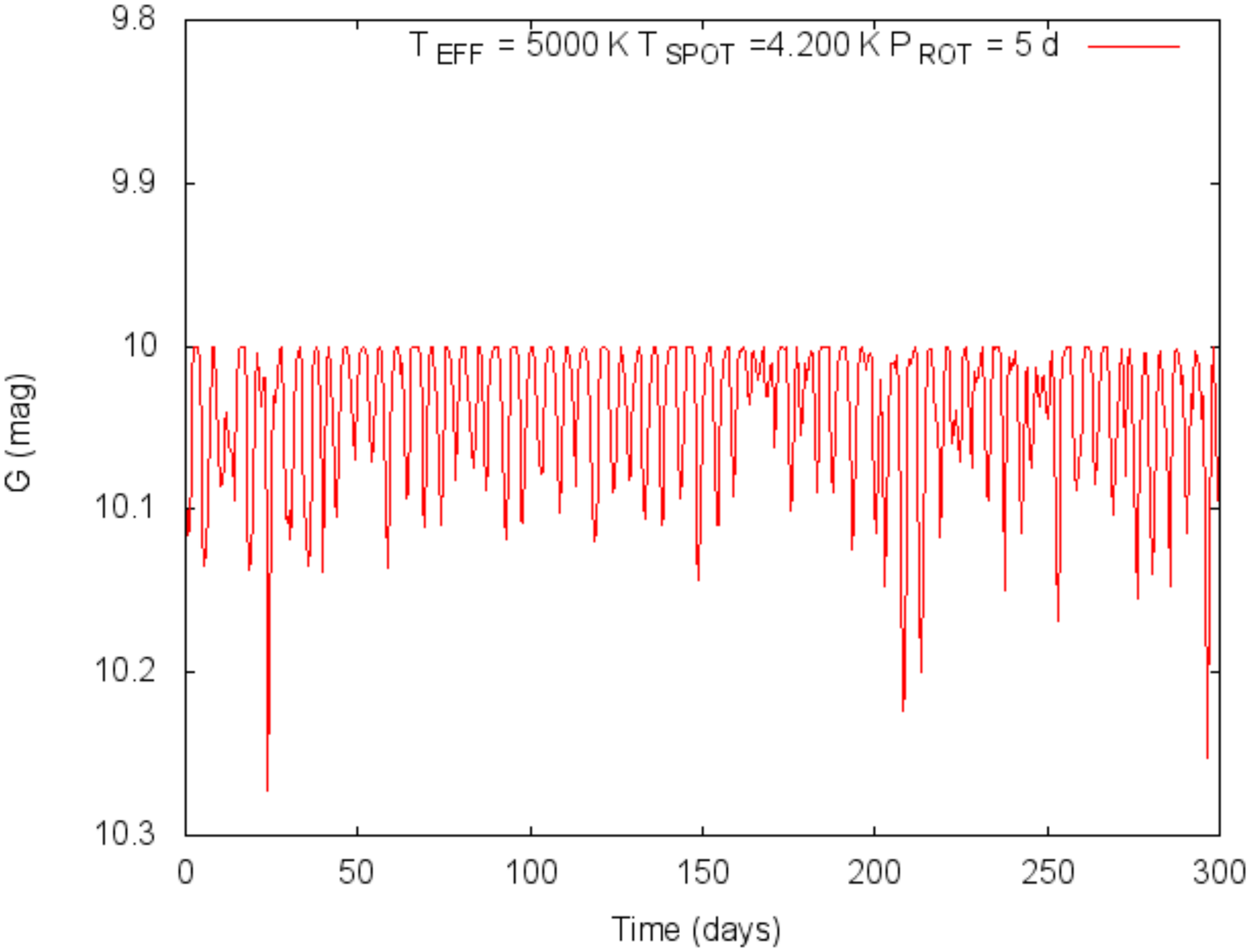}
\caption{Simulated light curve for a solar-like star with an effective temperature \teff = 5000 K, a rotation period $P$= 5\,d, and an average  $T_{\rm spot}$ = 4200\,K.
 }
\label{fig:simulatedcurve}
\end{center}
\end{figure}

These effects are illustrated in Fig.\,\ref{fig:simulatedcurve}, where we show the expected $G$ magnitude variations for a solar-like star with an effective temperature \teff=5000\,K, a rotation period $P$=5\,d, and an average $T_{\rm spot}$=4200\,K, according to the simulation code described in \cite{2012MNRAS.421.2774D}.
The evolution of the surface magnetic field produces changes in area, temperature, and location of the active region, which gradually distort the signal produced by the rotational modulation.

The segmentation of long-term time series also allows detecting period variations induced by the spot latitude migration combined with the stellar surface differential rotation (SDR).
When a star is characterised by SDR and by a solar-like cycle, where the spots  gradually migrate from intermediate latitudes to the equator, the detected rotation period changes slightly according to the cycle phase at which the star is observed \citep[see e.g.][]{2016A&A...591A..43D}.

The advantage of working with a more coherent pattern of flux variations in segments than in the whole time series comes at a price, as the segments contain fewer samplings and the period extracted from them may consequently have a lower statistical significance. 
Therefore, a compromise has to be found between the length of the segments  and the number of points in the segments.
\cite{2016A&A...591A..43D}, for instance, segmented the long-term photometric time series of a sample of young solar-like stars into 50\,d long sub-series by means of a sliding-window algorithm.  
Unfortunately, except for the observations of the ecliptic poles, this cannot be applied in general to the \gaia\ time series as many segments would have only two samplings $\simeq$106\,min apart, corresponding to two consecutive transits on the two telescope's FoVs \citep[see e.g.][]{2005MNRAS.361.1136E,2012MNRAS.421.2774D}.
In the nominal scanning law (NSL) observations\footnote{that
is, excluding the ecliptic poles scanning law (EPSL) that is
applicable to the initial observation phase}, segments 50\,d long would comprise $\sim$10--45 samplings only at ecliptic latitudes close to $\beta=\pm45\degree$. 

We therefore  modified the \cite{2016A&A...591A..43D} procedure into an adaptive segmentation algorithm that  takes  the peculiar features of the \gaia\ NSL into account.
In this new segmentation algorithm, the length of the sliding window is not fixed, but changes according to the sampling of each time series. 
The algorithm first searches groups of consecutive samplings separated by less than 6\,h.
Each group is identified by the set $\{t_{i_{\rm start}}, t_{i_{\rm end}}, N_i\}$, the observation time of the first point, the observation time of the last point, and the total number of observations per group.
Then all possible segments with extremities $t_{i_{\rm start}}$ and $t_{j_{\rm end}}$ with $j \ge i$, that is, all segments starting from the first point of a group to the final point of the same or any other subsequent group, are considered, and those satisfying the conditions 
\begin{eqnarray}
\Delta T &=& t_{j_{\rm end}} -t_{i_{\rm start}} \leq 120\,{\rm d} \label{condition1} \\ 
N_i + N_{i+1} + ... N_j &\geq& 12 \label{condition2} .
\end{eqnarray}
are selected for subsequent processing.
Nested segments are discarded, in the sense that if one segment satisfying the above conditions is entirely included in a longer segment, only the longer segment is considered.
In this way, the window length is set so that each segment covers an interval not exceeding 120\,d and includes at least 12 samplings.

The upper limit of the segment length (Eq.\,\ref{condition1}) may in some cases hamper a period detection in young stars, but this limit arises from an unavoidable compromise between the typical timescale of active region evolution and the characteristics of the \gaia\ sampling. 
At any rate, possible spurious periods that may results from active region evolution with timescales shorter than the segment length are filtered out, as discussed in Sect.\,\ref{finalselection}.

\subsection{Magnitude--colour variation correlation and outliers}
\label{sec:LinearCorrelation}

Magnitude and colour variations induced by rotation are linearly correlated, insofar as surface inhomogeneities maintain approximately the same location, area, and temperature, and their surface distribution is not uniform. 

This is confirmed by observations \citep{2006A&A...447..293M} and can be adopted as a sufficient, but not necessary, condition for rotational modulation detection in late-type stars.
The evolution of active regions may gradually change the characteristics of such a correlation so that when an isolated sampling is superimposed  on a group of close samplings at a distant epoch, it may appear as an outlier.
More significantly, flares may also appear as outliers in such a correlation, since they can be viewed as sudden changes of the surface inhomogeneity configuration produced by magnetic energy dissipation with consequent transient plasma heating.

Outliers are identified using the Theil-Sen robust linear regression between $G$ and \bmr\ \citep[see][for details]{theilsen1,theilsen2}.
Briefly, this algorithm computes the slopes and the intercepts of the straight lines passing through each couple of data points  ($G$, \bmr) measured in the segment.
The median slope and the corresponding intercept are taken as a first estimate of slope and intercept, and outliers are identified as points that are significantly distant from the straight line that best fits the data.
Furthermore, since flares can produce significant blueward changes in \bmr\ without greatly affecting the $G$ magnitude\footnote{The BP-bandpass is the most sensitive to flare events, while their effects are diluted in the $G$-band, which makes the colour \bmr\ more sensitive to flares than $G$.}, we add transits with 
\begin{equation}
\label{out}
(G_{\rm BP}-G_{\rm RP})_i < \overline{(G_{\rm BP}-G_{\rm RP})}- 5\sigma_{(G_{\rm BP}-G_{\rm RP})}
\end{equation}
to the list of outliers.
The identification of flare events amongst the outliers is discussed in  Distefano et al. (in prep.).
After the outliers are removed, the final slope and intercept are computed using a simple linear regression.

The Pearson correlation coefficient $r$ between the $G$ and \bmr\ variations is also computed.
The closer $r$ is to $\pm1,$ the higher the probability that rotational modulation is occurring, but period search (Sect.\,\ref{sec:PeriodSearch}) is carried out in all cases.

\subsection{Rotation period}
\label{sec:PeriodSearch}

Period search is performed using the generalised Lomb-Scargle periodogram method as implemented by \cite{2009A&A...496..577Z}.
This was previously adopted in the test analysis of \gaia\ DR1 data \citep{2017arXiv170203295E} and is best suited for the \gaia\ case \citep{2012MNRAS.421.2774D}.

The search is carried out on each segment and on the whole 22-month baseline.
Calling $T$ the time interval spanned by the segment (or by the whole time series), the frequency range in which the search is performed is
\begin{equation}
\left(\frac{2}{T}\,, ~3.2 \right) {\rm d}^{-1}
,\end{equation}
with a frequency step $(10 T)^{-1}$~d$^{-1}$.
The upper limit 3.2\,d$^{-1}$ is chosen in order to avoid the aliases corresponding to the 6-hour rotation period of the satellite.

The period with the highest power in the periodogram is selected, and the false-alarm probability (FAP), indicating the probability that the detected period is due to just noise, is associated with it.
In principle, methods based on Monte Carlo simulations would give the most reliable estimate of the FAP associated with a given period \citep[see e.g.][]{2012MNRAS.421.2774D}, but they are computationally expensive and prohibitive given the size of the \gaia\ sample.
\cite{2015MNRAS.450.2052S} demonstrated that the Baluev formulation \citep{2008MNRAS.385.1279B} is computational inexpensive and quite effective also in presence of strong aliases, and therefore this is adopted for the case at hand.

The choice of the FAP threshold has to balance the rates of false positives and true negatives.
For DR2, valid periods are chosen as those with FAP$\le$0.05, which corresponds to a false-detection rate of about $5\%$ \citep{2015MNRAS.450.2052S}\footnote{Also see \cite{2015MNRAS.450.2052S} for details on how the false alarm and correct detection rate depends on ecliptic latitude and on the number of observations.}

When a significant period $P$ is detected, a sinusoidal model 
\begin{equation}
\label{eq:model}
G(t)= A + B \sin\left(\frac{2\pi t}{P}\right) + C \cos\left(\frac{2\pi t}{P}\right) 
\end{equation}
is fitted to the time-series segment using the Levenberg-Marquardt method \citep{Levenberg_1944,Marquardt_1963}.

Periods inferred in different segments of the same source are finally combined using the mode as statistical estimator. 
For each source, a double loop is performed on the detected periods, counting for each period $P_i$, the number $N_i$ of periods that differ by less than 20\% \citep[which includes both uncertainties and the effects of differential rotation, see e.g.][]{2016A&A...591A..43D} from $P_i$.
We then consider the groups of similar periods $\{P_i\}_j$ (i.e. periods similar within the same group and different from one group to another considering the 20\% tolerance) with the highest frequency $N_{\rm max}$ and select the group with the lowest $\langle {\rm FAP} \rangle$.
The final period is given as the average of the periods in the selected group $P = \langle P_i \rangle$.

\subsection{Magnetic activity}
\label{AI}

The amplitude of the rotational modulation is related to the non-axisymmetric part of the active region distribution, which is a complex function of the projected area and the contrast in temperature between active regions and the unspotted stellar surface \citep[see e.g.][]{2006AN....327...21L,2012MNRAS.421.2774D}. 
The area and temperature of active regions, in turn, depend on the strength and topology of the surface magnetic field, which makes the amplitude of the rotational modulation a useful proxy for studying the stellar magnetic activity \citep[see e.g.][]{2000A&A...358..624R,2015A&A...583A.134F,2016A&A...588A..38L}.

Given the sparseness of the \gaia\ sampling and the presence of outliers (flares), we derive two quantities related to the rotational modulation amplitude, to be compared amongst each other for filtering the final results.
These are the differences between the 95$^{\rm th}$ ($G_{\rm 95th}$) and 5$^{\rm th}$ ($G_{\rm 5th}$) percentiles of the G magnitudes measured in the segment,
\begin{equation}
\label{activity_index}
A_{\rm per}=G_{\rm 95th} - G_{\rm 5th}
,\end{equation} 
and the amplitude derived from the sinusoidal fitting of the light curve (Eq.\,\ref{eq:model})
\begin{equation}
\label{afit}
A_{\rm fit}=2\sqrt{B^2 +C^2} .
\end{equation}
Both quantities give a measurement of the modulation amplitude that takes the presence of outliers into account, but they can be quite different when the light curve is not sampled with sufficient uniformity or when the sinusoidal model is a very poor representation of the data.
This is used in the final candidate filtering as described in Sect.\,\ref{finalselection}.

In DR2, an estimate of the unspotted magnitude is given as the brightest magnitude $G$ observed in each segment.
A more robust estimate, however, can be obtained from the Eq.\,(\refeq{eq:model}) fit coefficients as
\begin{equation} 
G_{\rm unspot} = A - \sqrt{B^2 + C^2}.
\end{equation}

\subsection{Final filtering}
\label{finalselection}

The algorithm described above found periodicity in some $7 \times 10^5$ sources in the Hertzsprung-Russel diagram (HRD) region described in Sect.\,\ref{sec:InputData}.
These may include spurious detections derived from the sampling incompleteness as well as other classes of periodic variability; eclipsing binaries and ellipsoidal variables are the only other expected classes.
Limiting the analysis to objects with a relative parallax uncertainty better than 20\% (Sect.\,\ref{sec:InputData}) makes the expected contamination from pulsating variables negligible, as only outlier in parallax would fall in the selected sample.\footnote{Cepheids, RR\,Lyrae, $\gamma$\,Doradus, etc. have much larger absolute magnitude $M_G$ than our selected sample.}
To minimise contamination as much as possible, we impose constraints on phase and amplitude in the period folded light curves without assuming
a strictly sinusoidal variation.

The constraints on phase are based on two parameters: the phase coverage (${\rm PC}$), and the maximum phase gap (${\rm MPG}$).
Data in a given segment are folded using the period detected in that segment and binned in equally spaced phase intervals\footnote{In DR2, each segment is divided into ten phase bins.}. 
The number of bins containing at least one data point is then divided by the total number of bins to obtain the phase coverage, which is therefore a real number in the $[0,1]$ interval.
A uniform phase coverage produces ${\rm PC} \rightarrow 1$, while a very concentrated phase coverage gives ${\rm PC} \rightarrow 0$.
Accepting only sources with ${\rm PC} \simeq 1$ would leave out many sources with reliable period estimates, but extending the accepted sample to ${\rm PC} \apprge 0.5,  $ for instance, would include some extreme cases with unacceptably large phase gaps, as the sampled phases can be concentrated in only half of the period-folded time series, for example.
In order to balance these two aspects, the ${\rm MPG}$ parameter,
\begin{equation}
\label{mpg}
{\rm MPG}= \max(\Delta\phi_{i,j})
,\end{equation}
where 
\begin{equation}
\label{phasedifference}
\Delta\phi_{i,j}=\phi_{i} -\phi{_j}
,\end{equation}
with $\phi_i$ and $\phi_j$ the phases of the $i^{\rm th}$ and $j^{\rm th}$ phase in the period-folded segment, is also taken into account.
In DR2, accepted sources are those satisfying the two requirements
\begin{eqnarray}
{\rm PC} &\ge & 0.4 \\
{\rm MPG} & \le & 0.3
\end{eqnarray}
in at least one segment, which ensures a good phase coverage with acceptable gaps.

The constraints on amplitude are based on the $(A_{\rm fit}/A_{\rm per})$ ratio and on the expected maximum amplitude.
In order to filter out light curves that are far from sinusoidal and/or have poor sampling, the $(A_{\rm fit}/A_{\rm per})$ distribution is cut at the 5$^{\rm th}$ and 95$^{\rm th}$ percentile, which corresponds at selecting sources with
\begin{equation}
\label{filter}
0.5 \le \left(\frac{A_{\rm fit}}{A_{\rm per}} \right) \le 1.6 \\
\end{equation}
only.
Regarding the expected amplitudes, the analysis of $\approx$30\,000 main-sequence stars observed by {\it Kepler} \citep{2014ApJS..211...24M} places them in the 0.006-0.025 mag range (white light).
\cite{2011A&A...532A..10M} found a $\Delta {\rm V} \approx 0.02-0.5$ rotational modulation range in stars belonging to the young loose associations $\epsilon$\,Cha and $\eta$\,Cha. 
\cite{1997A&AS..125...11S} reported an amplitude $\Delta {\rm V} = 0.65$ mag for the T Tauri star V410 Tau and an amplitude $\Delta \rm V~ = ~0.48 \rm ~mag$ for the giant star XX~Tri.
Based on previous knowledge, therefore, we conservatively filter out light curves with amplitude larger than 1 mag, considering it unlikely that such large variations can be due to just rotational modulation induced by surface inhomogeneities.

A further external filter is applied {\it \textup{a posteriori}} by cross-matching the final BY\,Dra sample with all other \gaia\ variability classes whose analysis can exploit more regular and stable light curves, and therefore their classification is considered more reliable than that of BY\,Dra.
The comparison with the still unpublished eclipsing binaries is particularly interesting because it gives an order of magnitude of the remaining eclipsing binaries contamination after filtering. 
Such a comparison resulted in $\approx 100$ sources in common between the BY\,Dra and the eclipsing binary classifications.

\begin{figure}[ht]
  \begin{center}
     \includegraphics[width=9cm]{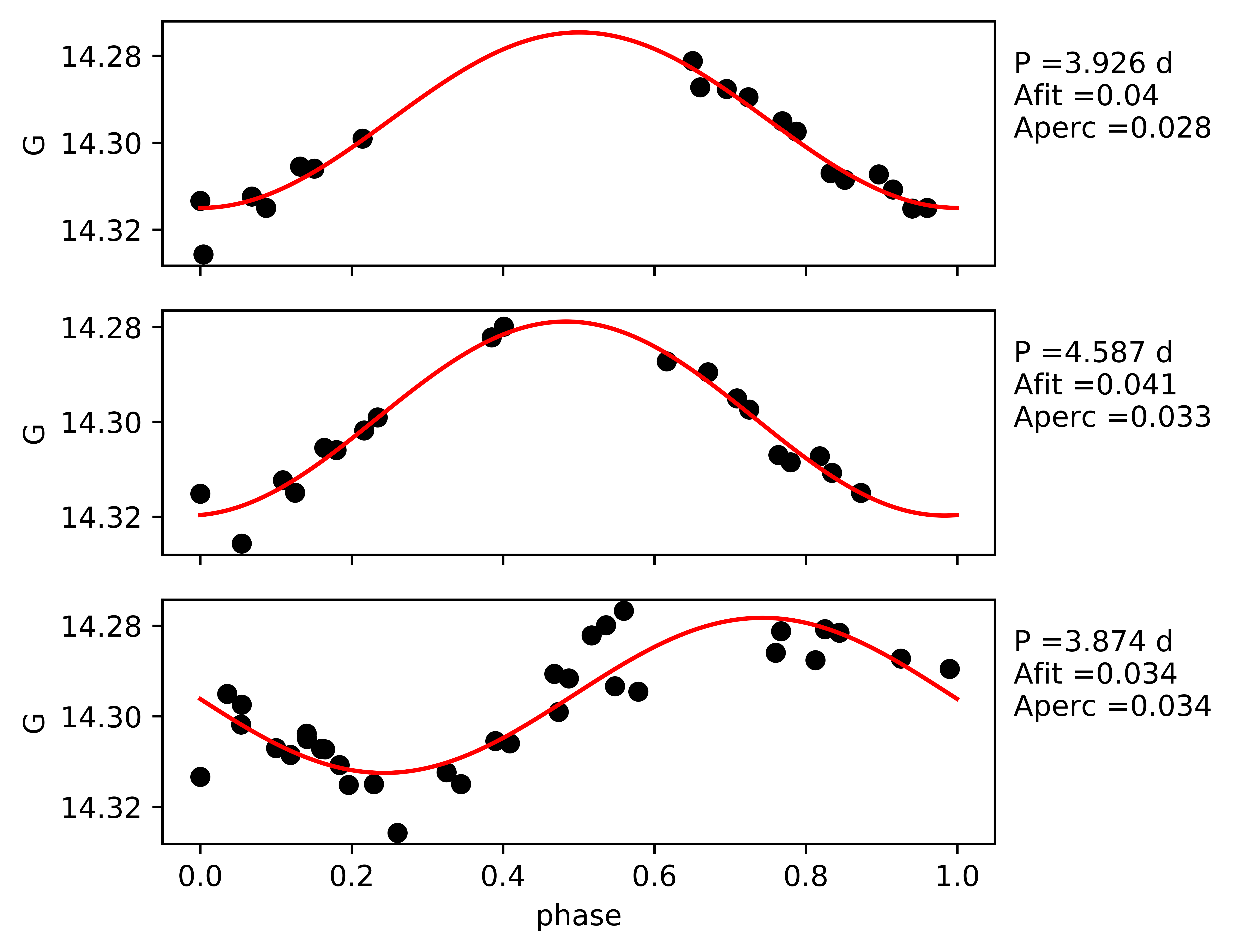}
      \caption{Period-folded light curves for the \gaia\ DR2 source id 894812322514469120. Three similar significant periods were found in two segments and in the whole time series with similar $A_{\rm fit}$ and $A_{\rm per}$.}
     \label{fig:goodexample}
  \end{center}
\end{figure}

An illustrative example of a final BY\,Dra candidate is shown in Fig.\,\ref{fig:goodexample} (\gaia\ DR2 source id 894812322514469120).
In this case, periods of $\approx 3.9 - 4.6$\,d are found in three segments with $A_{\rm fit}/A_{\rm per} \approx 1$ and a quasi-sinusoidal shape in all of them. 

For each source that passed the final filtering, the whole solution is contained in \gaia\ DR2.
This includes the source's time-series, the start and end observation times defining each segment, the solution found in each segment ($P$, $A_{\rm per}$, and the coefficients in Eq.\,\ref{eq:model}) and the source's final period and amplitude derived from all segments.

\section{Results}
\label{sec:Results}

        \subsection{Coverage}
        \label{sec:Coverage}

\begin{figure*}
  \begin{center}
     \includegraphics[width=14cm]{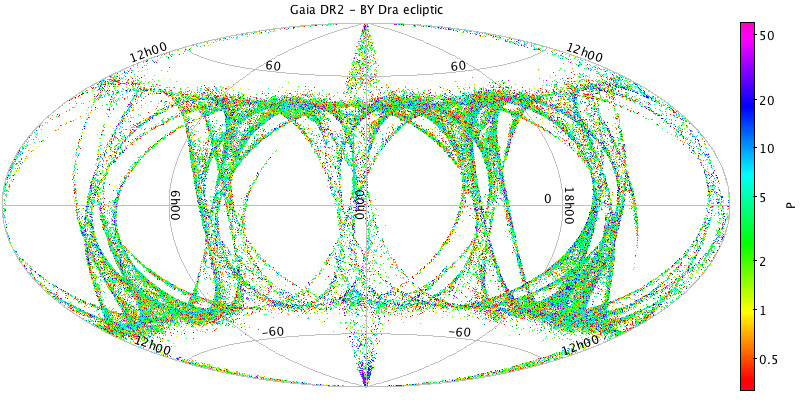}\\
     \includegraphics[width=14cm]{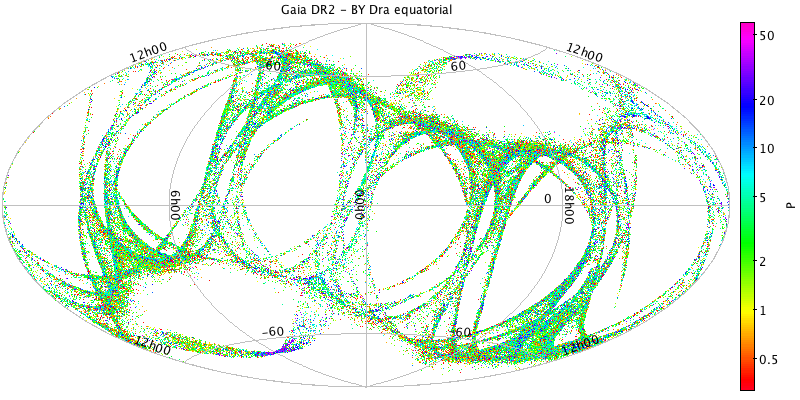}\\
     \includegraphics[width=14cm]{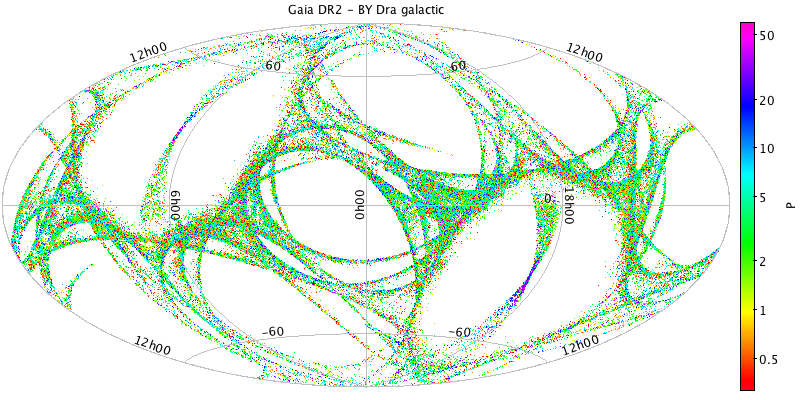}\\
      \caption{Sky distribution of \gaia\ DR2 BY\,Dra candidates. $P$ is given in day units.}
     \label{fig:map}
  \end{center}
\end{figure*}

The analysis of the first 22 months of \gaia\ observations with the method and filtering described in Sect.\,\ref{sec:Method} resulted in the identification of 147\,535 BY\,Dra variable candidates whose output parameters are included in \gaia\ DR2.
This sample covers 38\% of the whole sky when divided into bins of $\approx$0.84 square degrees \citep[level 6 HEALPix, see][]{2005ApJ...622..759G}.
Fig.\,\ref{fig:map} shows the sky distribution of period measurements.
The distribution is restricted to regions that are scanned more frequently in the 22-month baseline.
The filtering described in Sect.\,\ref{sec:Method} discards poorly sampled period-folded time series, so that the final sample has a range of 18 to 236 observations, with a median of 43, a mean of 45.65, a 5$^{\rm th}$ percentile of 28, and a 95$^{\rm th}$ percentile of 72 observations per target.

\begin{figure*}[ht]
  \begin{center}
     \includegraphics[width=8cm]{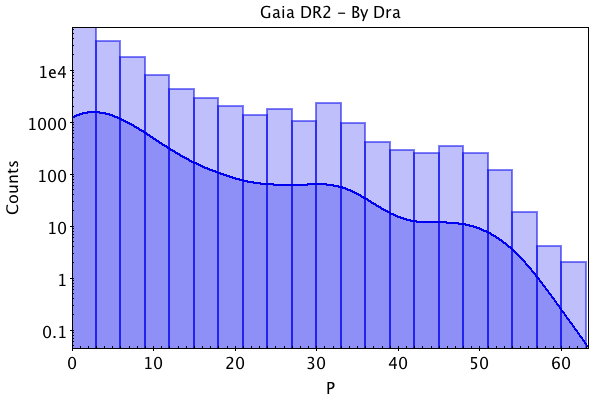}
     \includegraphics[width=8cm]{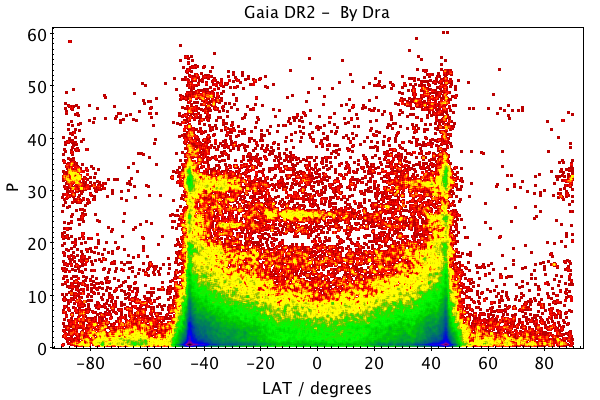}
      \caption{Rotation period distribution of \gaia\ DR2 data for BY\,Dra stars ($P$ is given in day units). A 3\,d binning $P$ histogram is shown in the left panel, together with the Gaussian kernel density estimate with a 3\,d width and 3-sigma truncation. The $P$ -- ecliptic latitude scatter plot (right panel) shows the higher number of period measurements at $\pm 45\degree$ ecliptic latitudes, particularly regarding longer periods. The colour code (rainbow) is an indication of the relative density of data points (red for lower density, and blue for higher density). Both panels give an indication of the moderate degree of aliasing, the most evident being at $\approx$\,23.5, 25.3, and 31.6\,d. }
     \label{fig:Pdistr}
  \end{center}
\end{figure*}

As anticipated by \cite{2012MNRAS.421.2774D}, the \gaia\ frequency sensitivity favours the detection of shorter periods ($\apprle 10$\,d), as shown in Fig\,\ref{fig:Pdistr}.
As expected, detection of longer periods is concentrated around ecliptic latitudes $\pm 45\degree$.
Fig.\,\ref{fig:Pdistr} also shows the presence of some moderate aliases, the more perceivable lie at $\approx$\,23.5, 25.3, and 31.6\,d.
We aim at some deeper understanding of these features in future analyses.

        \subsection{Completeness}
        \label{sec:Completeness}

Estimating the completeness of the BY Dra sample in \gaia\ DR2 is hampered by the lack of a well-established model for the expected number in the Galaxy.
It is expected that all stars with a convective envelope develop surface inhomogeneities, although in some cases these may not produce an observable rotational modulation. 
Detections of new low-mass dwarfs displaying rotational modulation is continuously increasing with the ever larger span and increasing sensitivity of modern surveys, none of which, however, has the full sky coverage capabilities of \gaia.

At this stage, it is possible to perform some meaningful comparison only with the HATNet \citep{2010MNRAS.408..475H} and K2 \citep{2016AJ....152..113R} surveys of the Pleiades, with which, nevertheless, the DR2 geq20 sky coverage still overlap only marginally (see Appendix\,\ref{completeness_appendix} for details). 
The \cite{2010MNRAS.408..475H} catalogue, in particular, is useful for a comparison with the Gaia data because of its uniform coverage of an approximately square sky region including the Pleiades cluster and an area close to it.
Furthermore, even though the Kepler (K2) pixel size is smaller than that of HATNet and Kepler can observe fainter stars than HATNet, the \cite{2010MNRAS.408..475H} catalogue contains a larger number of non-members in a smaller area, making our rough estimate easier and more realistic (see Appendix\,\ref{completeness_appendix}).
Assuming that the \cite{2010MNRAS.408..475H} catalogue lists all the BY Dra in its FoV down to $G\approx14.5$, we estimate that the completeness of the  BY~Dra sample is 14\% in the overlapping field.
Assuming that this value is uniform over the whole sky, we estimate a completeness upper limit of 5\%.
At the other extreme, we may assume that all low-mass dwarfs are BY~Dra variables.
Then, comparing with all stars observed by \gaia\ in the same sky region and magnitude range, we estimate a lower completeness limit of 0.7\%.

        \subsection{Contamination}
        \label{sec:Contamination}

An estimate of the contamination can be based on the type and amount of 
expected contaminants left after our pre-selection, based on stars around the main sequence, and our filtering (Sect.\,\ref{sec:Method}), which discards results with FAP$>$0.05 and light curves that
are far from being sinusoidal.
Neglecting outliers in parallax and the expected false-detection rate of about 5\%, the only remaining expected main contaminants are close grazing binaries with a period $\apprle 10$\,d, whose light curve is not far from being sinusoidal.
When the binary period is long, then the grazing eclipse covers a limited phase interval, and it does not look like rotation modulation.

We assume a upper limit period of 10\,d for this contaminating effect. 
The fraction of stars with such short-period companions is only about 2--5\% \citep{2010ApJS..190....1R},  out of which only 2--5\% are grazing binaries (at most), which means 0.04--0.25~\% of the stars, while we did not yet take into account that the secondaries are substantially smaller in most binaries. 
Even though we cannot consider the rotation modulation sample a random sub-set of G and later type stars (which is the assumption in above estimate), it seems unlikely that this type of contamination would exceed 1\%. The same conclusion is reached from our internal validation against known and newly identified (but not published) eclipsing binaries, based on which we estimate an upper limit of 0.5\% grazing binaries contamination.

\subsection{Validation of rotation periods}

\begin{figure*}
\begin{center}
\includegraphics[width=0.44\textwidth]{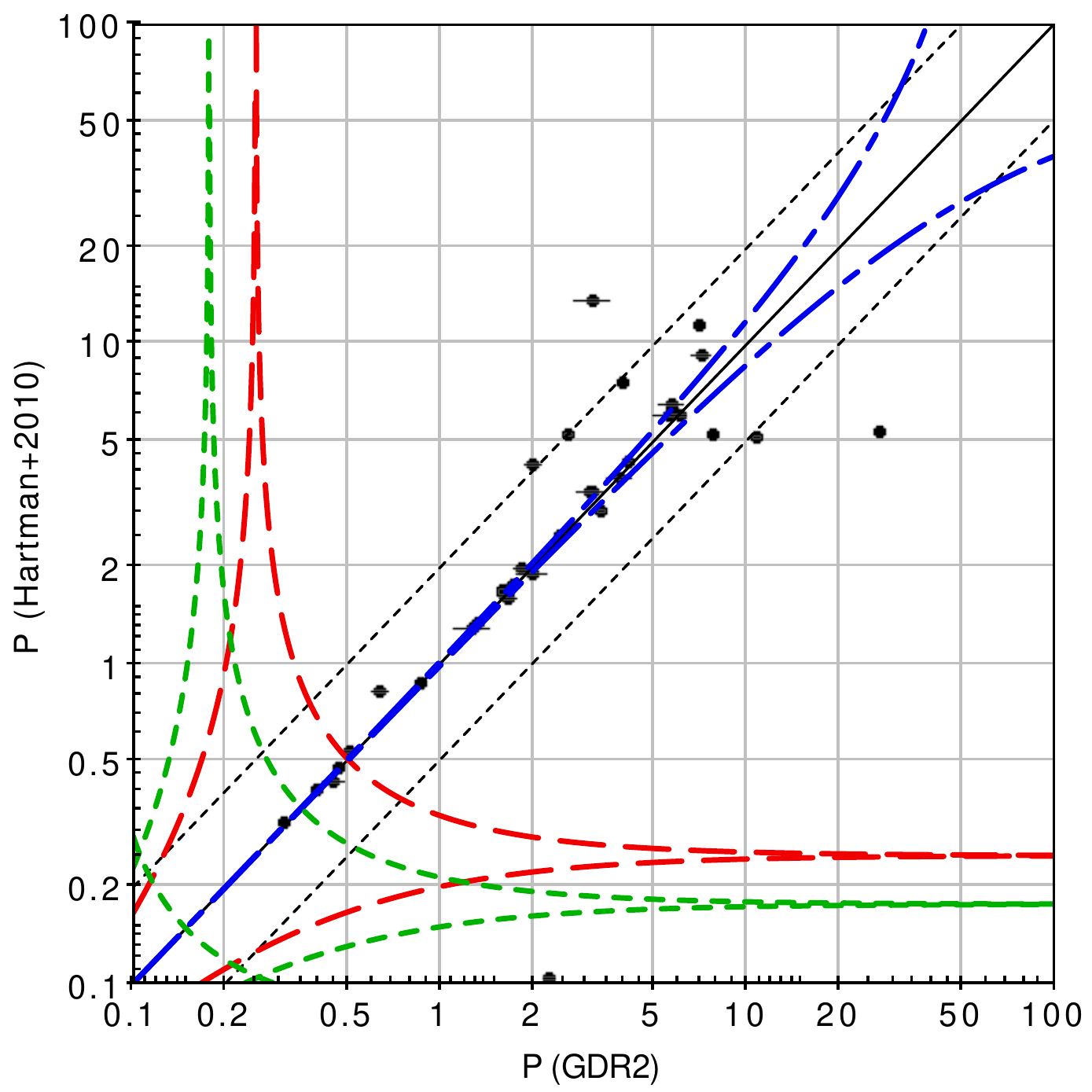}
\includegraphics[width=0.44\textwidth]{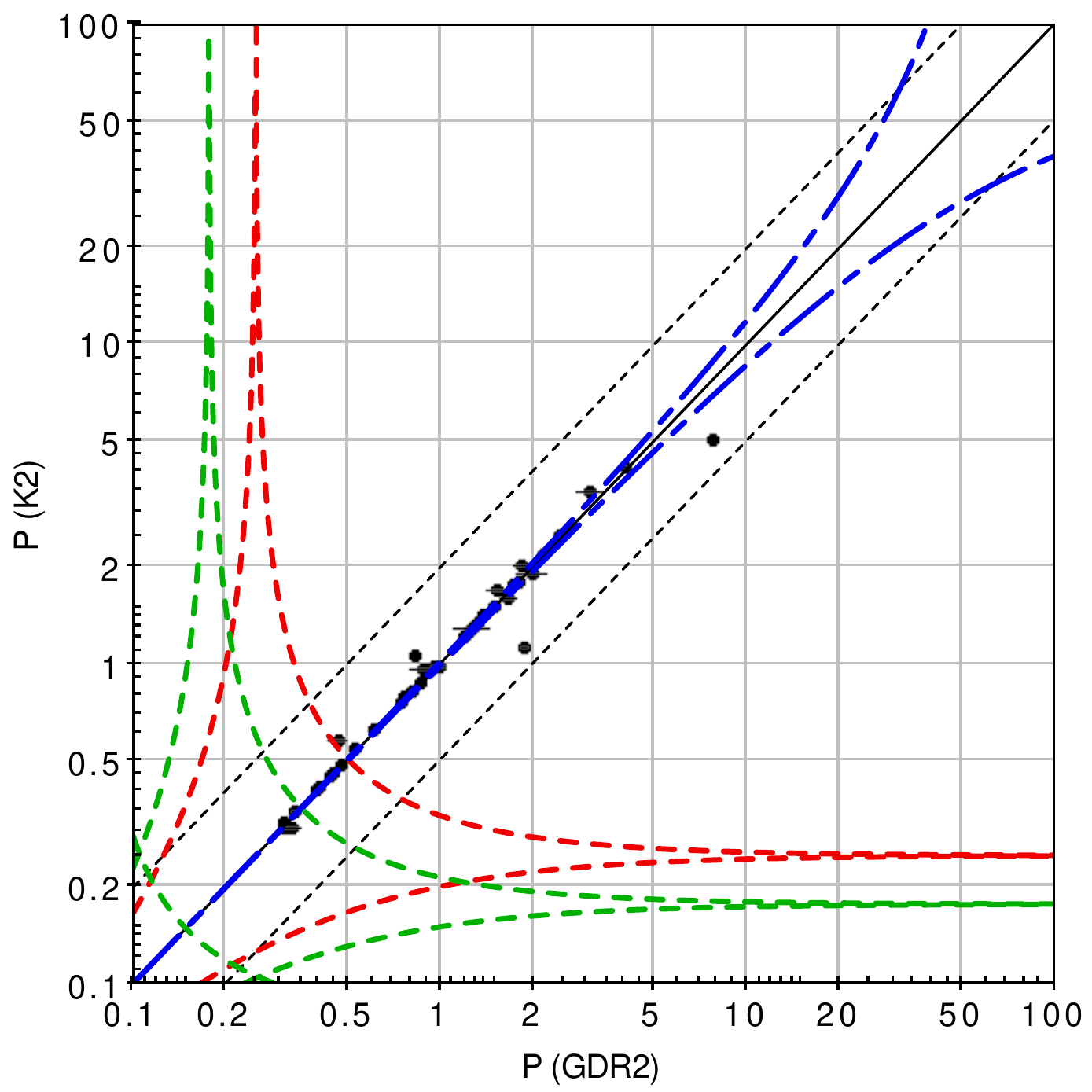}
\caption{Left panel: Period comparison with the \cite{2010MNRAS.408..475H} Pleiades data.
Black short dashed lines are the half and double period loci, 
red long dashed lines show the spacecraft rotation alias loci, 
blue dot-dashed represent the alias loci associated with Gaia spin axis precession,
and green short dashed lines show the alias associated with the time delay between the two FoV (basic angle).
Right panel: Period comparison with the K2 observations of the Pleiades \citep{2016AJ....152..113R}.
$P$ is given in day units.}
\label{fig:P_comparison}
\end{center}
\end{figure*}

Comparison with the rotational periods of stars in common with
the \cite{2010MNRAS.408..475H} and \cite{2016AJ....152..113R} dataset is shown in Fig.\,\ref{fig:P_comparison}.
There are 36 stars in common with \cite{2010MNRAS.408..475H}, 4 of which have a period close to the half or double period, and 25 agree to better than 10\%.
Of the 40 stars in common with \cite{2016AJ....152..113R}, 2 are close to the half-period, and 36 agree to better than 10\%.
We therefore conclude that the robustness and accuracy of the \gaia\ rotational period recovery is comparable to other data published in the literature.

\subsection{Period - colour and modulation amplitude distributions}

\begin{figure}
\begin{center}
\includegraphics[width=0.44\textwidth]{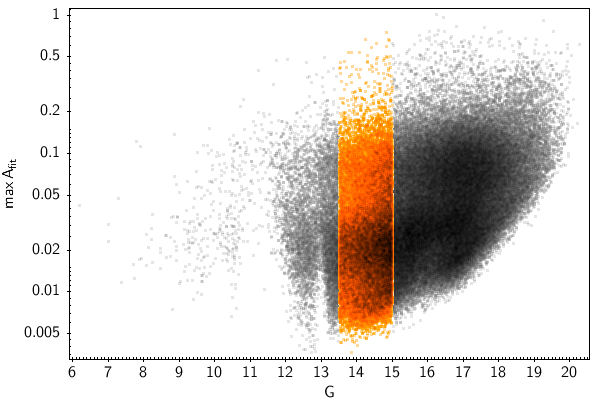}
\caption{Modulation amplitude (in mag.) dependence on apparent magnitude $G$. The period-colour diagram and amplitude distribution reported in this paper are built using only the subsample highlighted in orange.}
\label{fig:Ampl_G}
\end{center}
\end{figure}

The lowest values of the modulation amplitude have a dependence on the apparent $G$ magnitude that closely mimics the
dependence of the photometric sensitivity on $G$ (Fig.\,\ref{fig:Ampl_G}).
In order to simplify the description as best possible, that is,
to avoid complications arising from a $G$-dependent completeness level, we select a subsample in the range $13.5 \le G \le 15.0$ mag, for which the modulation amplitude sensitivity is approximately constant over $G$.

\begin{figure*}
\begin{center}
\includegraphics[width=0.40\textwidth]{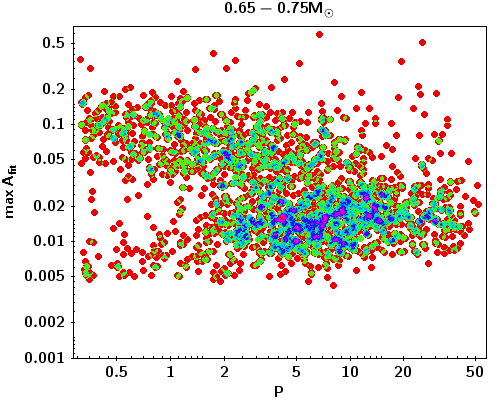}
\includegraphics[width=0.40\textwidth]{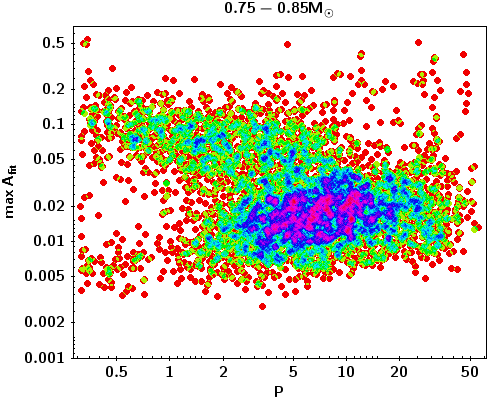}
\includegraphics[width=0.40\textwidth]{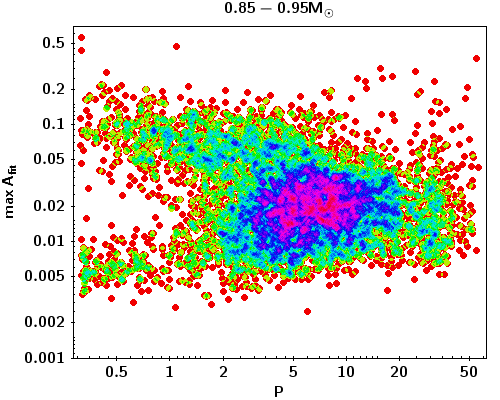}
\includegraphics[width=0.40\textwidth]{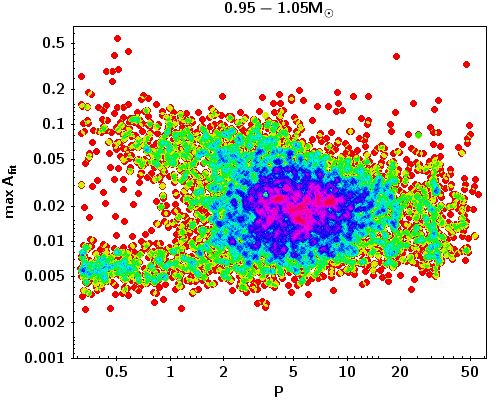}
\includegraphics[width=0.40\textwidth]{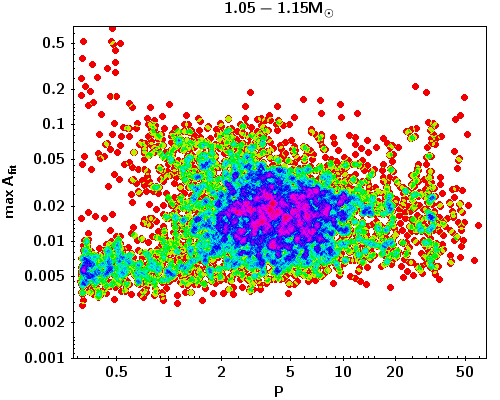}
\includegraphics[width=0.40\textwidth]{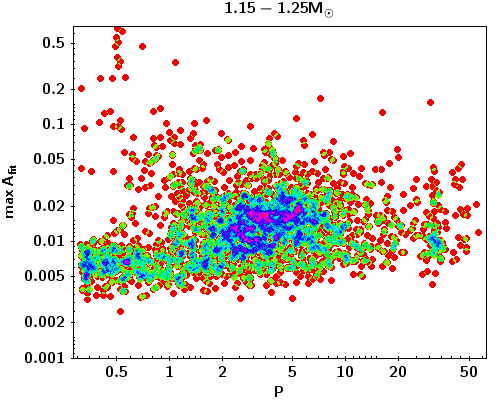}
\caption{Amplitude-period diagrams in bins of approximate mass showing a bimodality in modulation amplitude. The colour coding gives an indication of the relative density of the data points (red for lower-density, and purple for higher density).
Amplitude units are mag., and $P$ is given in day units.}
\label{fig:Ampl}
\end{center}
\end{figure*}

The selected subsample is further divided into stellar mass bins using $M_{G}$ as an approximate proxy, which is first converted into $M_{V}$ \citep[][and references therein]{2018arXiv180409368E} and then into mass using the relationships given in \cite{2010ApJ...721..675B}. 
In Fig.\,\ref{fig:Ampl} we show the distributions of the modulation amplitude\footnote{Each source is characterised by the maximum amplitude over segments in which the period search was successful.} over $P$ in $\Delta M \approx 0.10 M_\odot$ bins from 0.65 to 1.25 $M_\odot$.
The data show a clear bimodality with a boundary around $\max{A_{\rm fit}} \approx 0.04 - 0.05$.
For $P<2$\,d, there is also a clear gap between the low-amplitude and the high-amplitude population from $\approx$ 0.01 to 0.05 mag.
The high-amplitude population is more prominent at lower mass and is gradually depleted as we consider higher mass, until it almost disappears in the 1.15 - 1.25 $M_\odot$ bin.
Conversely, the low-amplitude population is very poorly populated for $M \apprle 0.85 M_\odot$ in the $P \apprle 2$\,d region, but this region is increasingly populated as we consider higher mass.

\begin{figure*}
\begin{center}
\includegraphics[width=0.80\textwidth]{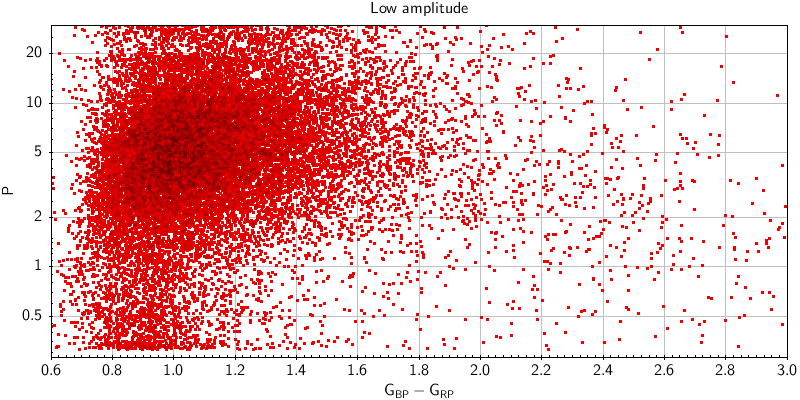}
\includegraphics[width=0.80\textwidth]{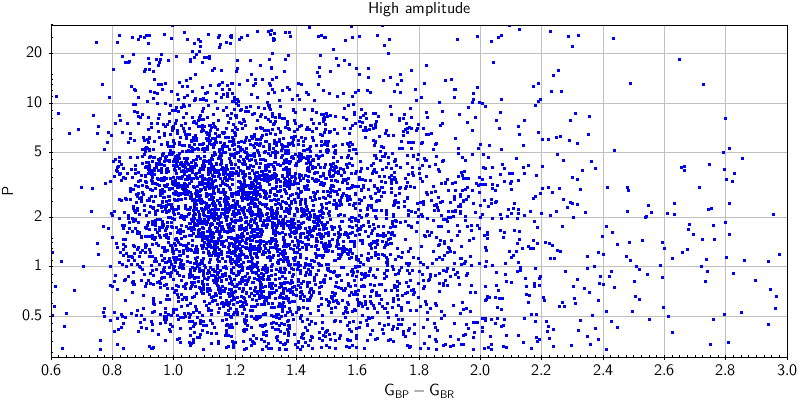}
\caption{Period-colour diagrams for the low-amplitude population ($\max{A_{\rm fit}} \le 0.04$, upper panel, and for the high-amplitude population $\max{A_{\rm fit}} > 0.04$, lower panel).
$P$ is given in day units.}
\label{fig:PC}
\end{center}
\end{figure*}

The low- and high-amplitude populations identified in the amplitude-period diagram also tend to occupy distinctive areas in the period-colour diagram.
The period-colour diagrams of these two populations (Fig.\,\ref{fig:PC}) show that the low-amplitude population tends to mostly occupy
the region where we expect to find a collection of Barnes' I-sequences \citep[or slow-rotator sequences,][]{2003ApJ...586..464B} at different ages, while the high-amplitude population tends to mostly occupy the region where we expect a collection of Barnes' C-sequences and the gap between the C- and the I-sequence.

We investigated possible instrumental or analysis artefacts that could spuriously produce these distributions, but we found no evidence of this. 
The bimodal distributions described above can indeed be an indication of two different modes of operation of the hydromagnetic dynamo in stars with the same mass, effective temperature, and rotation period. 
The Sun has time intervals of prolonged low activity, called grand minima, lasting for several decades. They are separated by intervals of normal activity characterised by the 11-year cycles and greater  variability amplitudes. 
Grand minima occur for about 17-20\% of the time \citep[e.g.][]{2007A&A...471..301U,2009A&A...496..577Z}.
In the case of more active stars, the possibility that the dynamo is operating in different regimes has also been suggested, for example, to account for the distributions of the rotation periods of late-type stars in open clusters \citep[e.g.][]{2003ApJ...586..464B,2014ApJ...789..101B}. 
Previous observations did not clearly detect the bimodality because of the lower photometric precision of ground-based photometry or a rather low percentage of low-amplitude fast-rotators in Kepler data \citep{2014ApJS..211...24M,2015A&A...583A..65R}. 

Contamination by grazing eclipsing binaries (Sect.\,\ref{sec:Contamination}) is expected to have negligible effects on the distributions described above.
Given the short periods where the bimodality is mostly concentrated, the contaminating eclipsing binaries are expected to have synchronised components. 
Eclipsing binaries with deep eclipses, larger than about 0.015 mag, for instance, are rejected by our filtering criteria, but grazing binaries or ellipsoidal variables with small amplitude ($<$ 0.01 mag) can escape filtering and might be misclassified as rotational variables, especially because the intrinsic variability of their components due to their magnetic activity effectively increases the level of noise in their light curves, hiding the true shape of the modulation.
However, since the expected percentage of such objects is well below 1\%, they have no significant effects on the distributions shown in Figs.\,\ref{fig:Ampl} and \ref{fig:PC}.

\section{Conclusions}
\label{sec:Conclusions}

We have presented the method we used to identify and analyse rotational modulation variables of the BY\,Dra class in the second \gaia\ data release.
This contains 147\,535 BY\,Dra candidates, together with their photometric time series and the essential parameters related to their flux modulation induced by surface inhomogeneities and rotation.
Although this sample represents only a few percent of all BY\,Dra that can be studied with \gaia, this is already by far the largest BY\,Dra catalogue available to date.
False positives are estimated to be of the order of 5\%.
The main contaminants are essentially limited to low-mass short-period grazing binaries or binary ellipsoidal variables whose light curves are not far from being sinusoidal.
Their expected fraction, however, is expected to be below 1\%, making their effects on the distributions over the parameters of interest negligible. 

The richness of the sample, the range of periods, and the photometric precision make it possible for the first time to unveil a clear bimodality in the amplitude-period diagrams binned by mass, with a gap between the two populations at $P<2$\,d.
The high-amplitude population is particularly evident at lower stellar mass ($\apprle 1 M_\odot$) and is gradually depleted at increasing mass.
Conversely, the low-amplitude population at $P<2$\,d is particularly rich at higher mass ($\apprge 1 M_\odot$) and is increasingly depleted at lower mass.
Furthermore, the low-amplitude population is mainly located in regions of the period-colour diagram resembling a collection of Barnes' I-sequences at different ages \citep{2003ApJ...586..464B}, while the high-amplitude population is mainly located in regions resembling C-sequences and the gaps between the two.
A deeper analysis would require the inclusion of stellar age and $G$-dependent completeness estimates, but this is beyond the scope of this paper.
However, the very existence of this bimodality is further confirmation of the existence of different modes of operation of the dynamo in stars with the same mass, effective temperature, and rotation period.

The possibility of distinguishing such new details in the sample distribution over the parameters of interest, which independently confirm a scenario that is in line with previous studies, is by itself an indication of the validity and quality of the BY\,Dra \gaia\ DR2 data.
Comparison of the \gaia\ rotation periods with literature data, although limited to a few tens of cases, confirms that the accuracy and robustness is at the same level as those of other surveys, including, notably, those conducted with Kepler. 

Overall, the preliminary data contained in \gaia\ DR2 illustrate the vast and unique information that the mission is going to provide on stellar rotation and magnetic activity, opening new and unique opportunities in our understanding of the evolution of stellar angular momentum and dynamo action.

\begin{acknowledgements}
This work presents results from the European Space Agency (ESA) space mission Gaia. 
Gaia data are being processed by the Gaia Data Processing and Analysis Consortium (DPAC). Funding for the DPAC is provided by national institutions, in particular the institutions participating in the Gaia MultiLateral Agreement (MLA). 
This work was supported by the Italian funding agencies Agenzia Spaziale Italiana (ASI) through grants I/037/08/0, I/058/10/0, 2014-025-R.0, and 2014- 025-R.1.2015 to INAF and contracts I/008/10/0 and 2013/030/I.0 to ALTEC S.p.A and Istituto Nazionale di Astrofisica (INAF)  (PI M.G. Lattanzi). 
For Switzerland this work was supported by the Swiss State Secretariat for Education, Research and Innovation through the ESA PRODEX program, the ``Mesures d'accompagnement'', the ``Activit\'es Nationales Compl\'ementaires'', the Swiss National Science Foundation, and the Early Postdoc. Mobility fellowship.
We acknowledge the use of TOPCAT \citep[\href{http://www.starlink.ac.uk/topcat/}{http://www.starlink.ac.uk/topcat/},][]{2005ASPC..347...29T}.
The Gaia mission website is \href{https://www.cosmos.esa.int/gaia}{https://www.cosmos.esa.int/gaia}. 
The Gaia Archive website is \href{http://gea.esac.esa.int/archive/}{http://gea.esac.esa.int/archive/}.
Thanks to Sydney Barnes, AIP Potsdam, for inspiring discussions. 
\end{acknowledgements}

\bibliographystyle{aa}
\bibliography{GaiaSLDR2}


\appendix

\section{Estimating completeness}
\label{completeness_appendix}

Limiting our considerations to pre- and main-sequence single stars or wide binaries, there are several factors to be considered:
\begin{itemize}
\item In the main sequence, the amplitude of the flux modulation decreases while the period increases with stellar age; rotational modulation is therefore increasingly difficult to detect as the star evolves.
\item Solar-like activity is characterised by magnetic cycles, so that  rotational modulation is more easily detectable at epochs close to a maximum of the cycle and may be completely undetectable at epochs close to a minimum. 
\item Stars seen pole-on or quasi-pole-on are expected to produce negligible rotational modulation.
\end{itemize}

Bearing these considerations in mind and in view of what we showed in Sect.\,\ref{sec:Coverage}, in order to estimate the completeness of the BY\,Dra sample in \gaia\ DR2, we start by defining a detection efficiency
\begin{equation}
Q = Q(\lambda,\beta,P,G,(G_{\rm BP}-G_{\rm RP})) ,
\label{eq:efficiency_def}
\end{equation}
as a function of the ecliptic coordinates\footnote{The Gaia scanning law is naturally more regular in the ecliptic reference system, and therefore a description of the dependence of the efficiency in recovery a given period on a given direction is simpler when using this reference system. See also Fig.\,\ref{fig:map}} $(\lambda,\beta)$, the period $P$, the apparent magnitude $G,$ and colour \bmr, such that
\begin{equation}
Q \equiv \frac{N}{\hat{N}} 
\label{eq:def_efficiency}
,\end{equation}
where $N=N(\lambda,\beta,P,G,(G_{\rm BP}-G_{\rm RP}))$ is the number density of detected BY\,Dra and $\hat{N}=\hat{N}(\lambda,\beta,P,G,(G_{\rm BP}-G_{\rm RP}))$ the true number density of BY\,Dra. 
Since we do not know $\hat{N}$, we make use of two estimates that may reasonably bracket the real value: at one extreme, we assume that all low-mass stars selected for the analysis in the HRD are BY\,Dra (say $\hat{N} \approx \hat{N}_0$), at the other, we assume that a reference survey (RS) has detected all BY\,Dra in the field, explored down to its completeness magnitude limit (e.g. $\hat{N} \approx \hat{N}_{\textrm{RS}}$). 

To a certain extent, the \gaia\ DR2 BY\,Dra sample overlaps in a region around the Pleiades cluster with the HATNet \citep{2010MNRAS.408..475H} and K2 \citep{2016AJ....152..113R} surveys.
The \cite{2010MNRAS.408..475H} catalogue is particularly useful for a comparison with the \gaia\ data because of its uniform coverage of an approximately square sky region including the Pleiades cluster and an area close to it.
Furthermore, this catalogue contains very many non-members in a smaller area than K2, which makes our estimate easier and more realistic.
We therefore take this survey as reference for our completeness estimate.
Moreover, since the coverage in $(\lambda, \beta)$ is still sparse, 
Eq.\,\eqref{eq:def_efficiency} is evaluated in the direction of the Pleiades and the value obtained taken as representative for the whole sky for the time being, deferring the full analysis implied by Eq.\,\eqref{eq:def_efficiency} to a later stage when a more uniform BY\,Dra sky coverage will become available with a more extended \gaia\ baseline.

\begin{figure*}
\begin{center}
\includegraphics[width=0.50\textwidth]{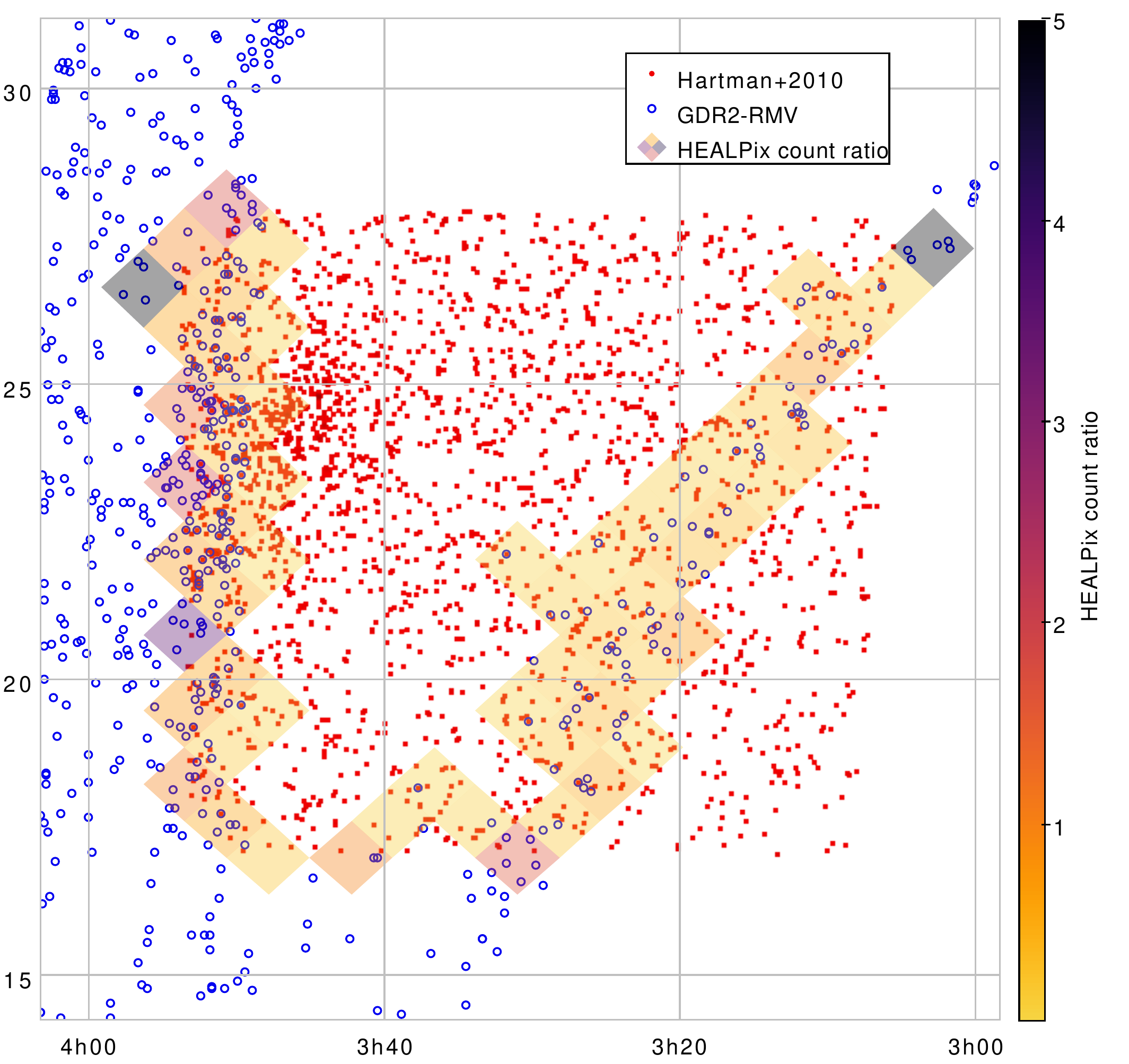}
\includegraphics[width=0.44\textwidth]{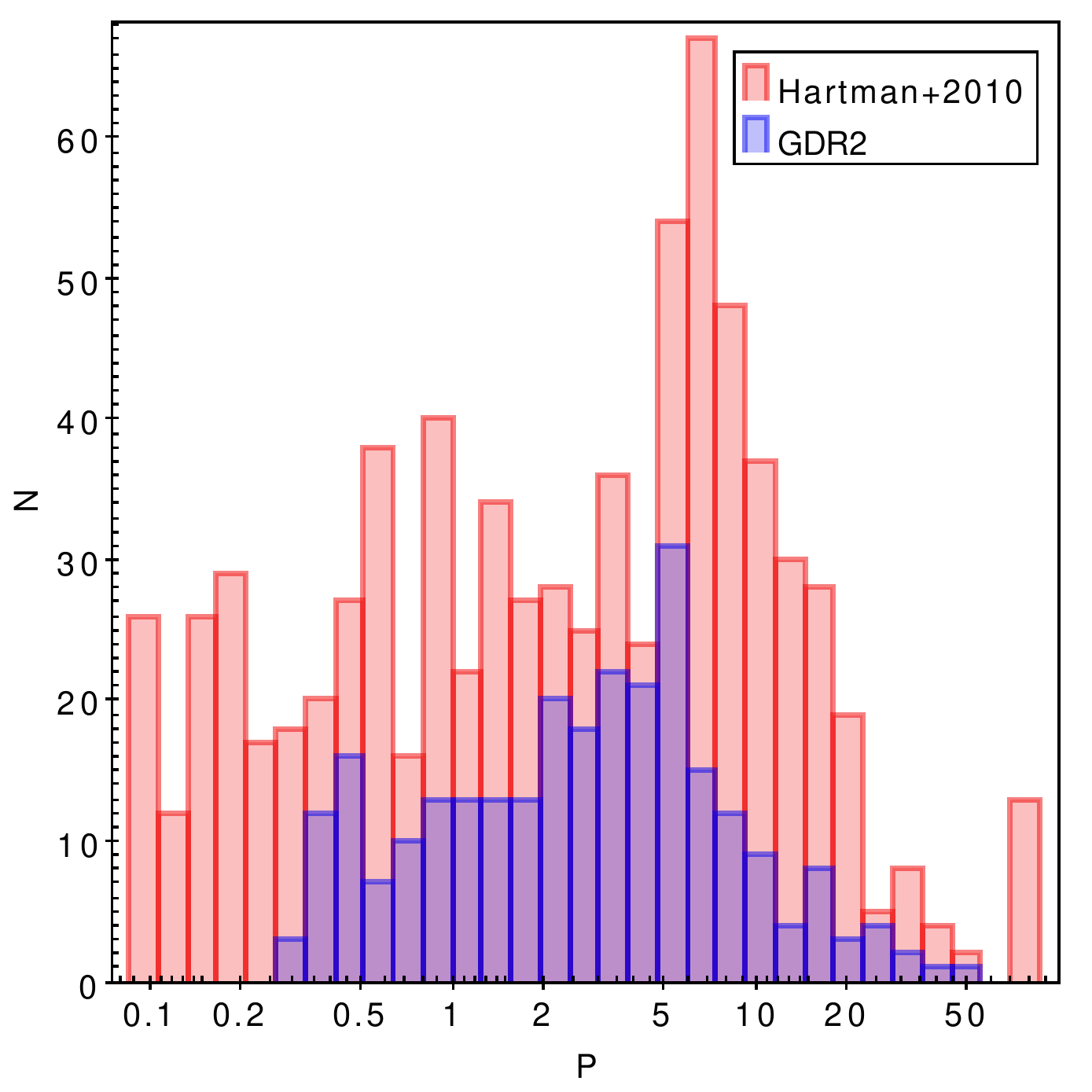}
\includegraphics[width=0.44\textwidth]{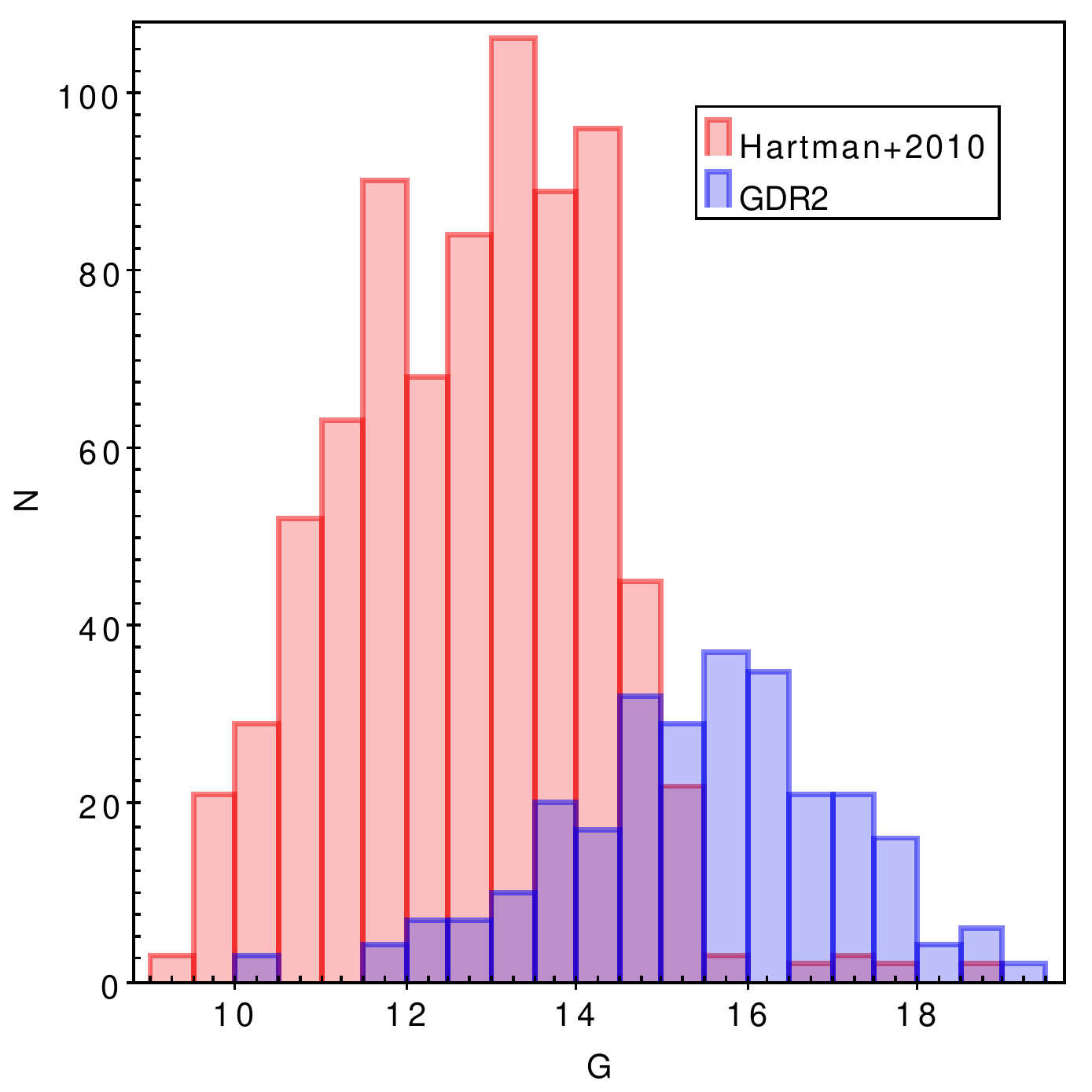}
\includegraphics[width=0.44\textwidth]{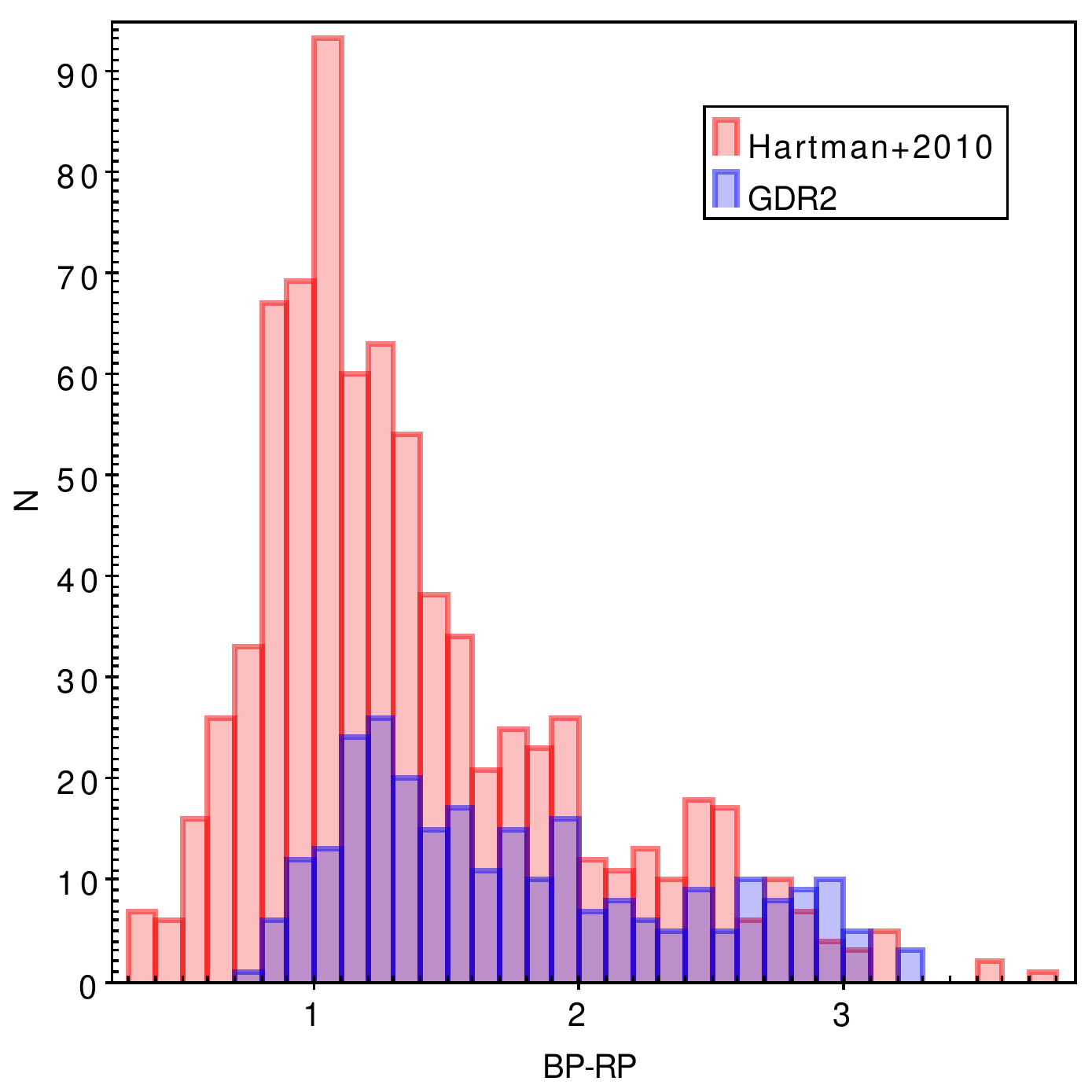}
\caption{Upper left panel: sky map of the \cite{2010MNRAS.408..475H} sample (red filled dots) with the \gaia\ rotational modulation variable sample (blue open dots) overlapped.
The \texttt{HEALPix} in which we have stars from both samples are outlined, and the colour code indicates the star number ratio in each \texttt{HEALPix}.
Upper right and bottom panels: Comparison of the $P$, $G$, and \bmr\ distributions in \gaia\ and the \cite{2010MNRAS.408..475H} sample in the overlapping \texttt{HEALPix}. 
$P$ is given in day units.}
\label{fig:Hartman_raw_comparison}
\end{center}
\end{figure*}

A sky map of the \cite{2010MNRAS.408..475H} data with the DR2 rotational modulation variable sample overlapped is shown in Fig.\,\ref{fig:Hartman_raw_comparison}.
In order to estimate the $N/\hat{N}_{\textrm{RS}}$ ratio, the whole sphere in equatorial coordinates has been divided into \texttt{HEALPix} pixels with resolution index 6, that is to say, a division of the whole sphere into $12 \times 6^2=49152$ pixels of size of 0.83929366 square degrees.
In Fig.\,\ref{fig:Hartman_raw_comparison}, \texttt{HEALPix} with stars belonging to both samples are outlined with a colour code indicating the integrated (in $P$, $G$, and \bmr) $N/\hat{N}_{\textrm{RS}}$ ratio in each \texttt{HEALPix}.
In this case, $N/\hat{N}_{\textrm{RS}}$  extends from 0.02 to 5, with a median of 0.29.
The extreme $N/\hat{N}_{\textrm{RS}}$ values correspond to the borders of either the \cite{2010MNRAS.408..475H} field or the \gaia\ strips, and therefore we take the median value as a representative estimate.

\begin{figure*}
\begin{center}
\includegraphics[width=0.50\textwidth]{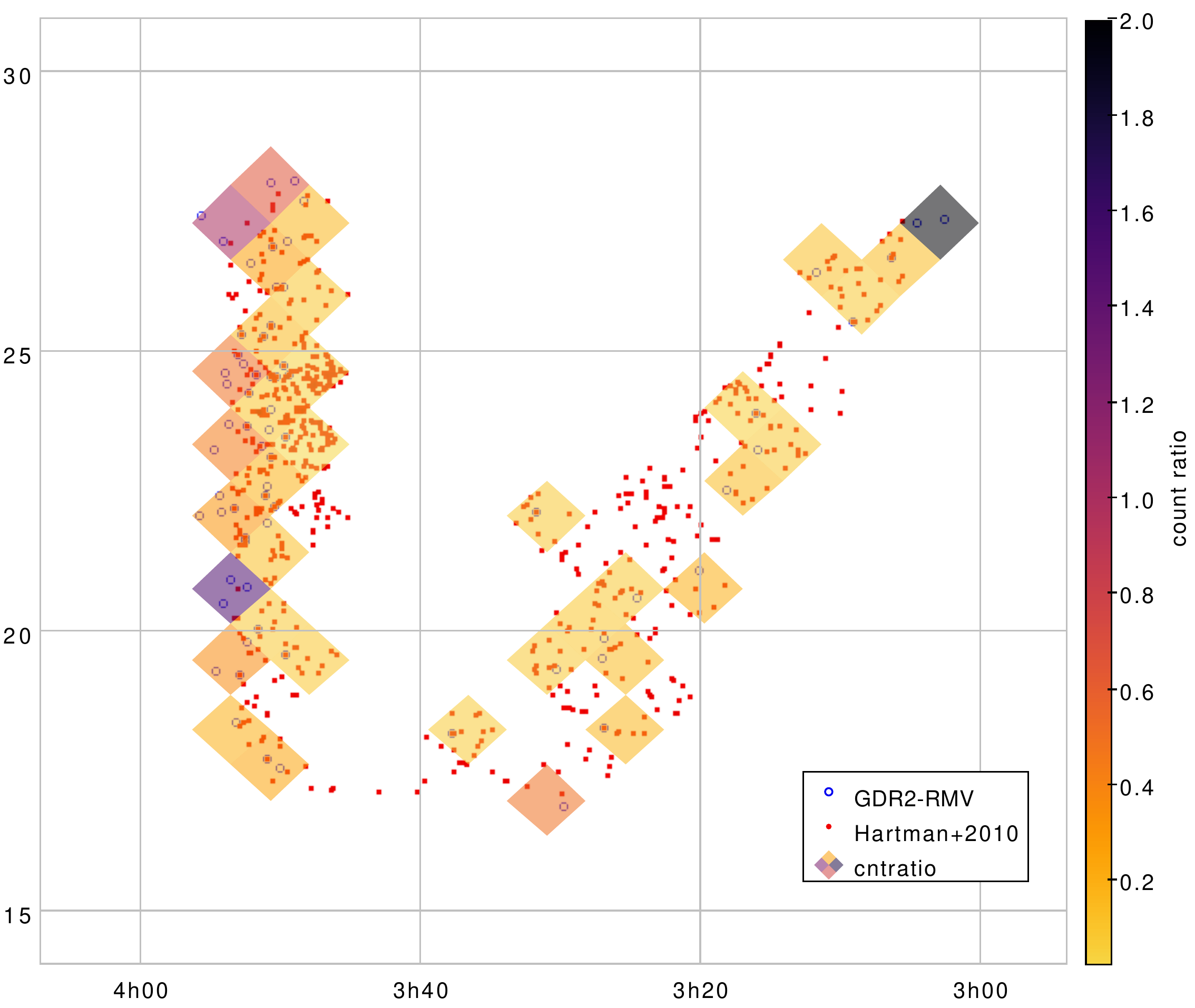}
\includegraphics[width=0.44\textwidth]{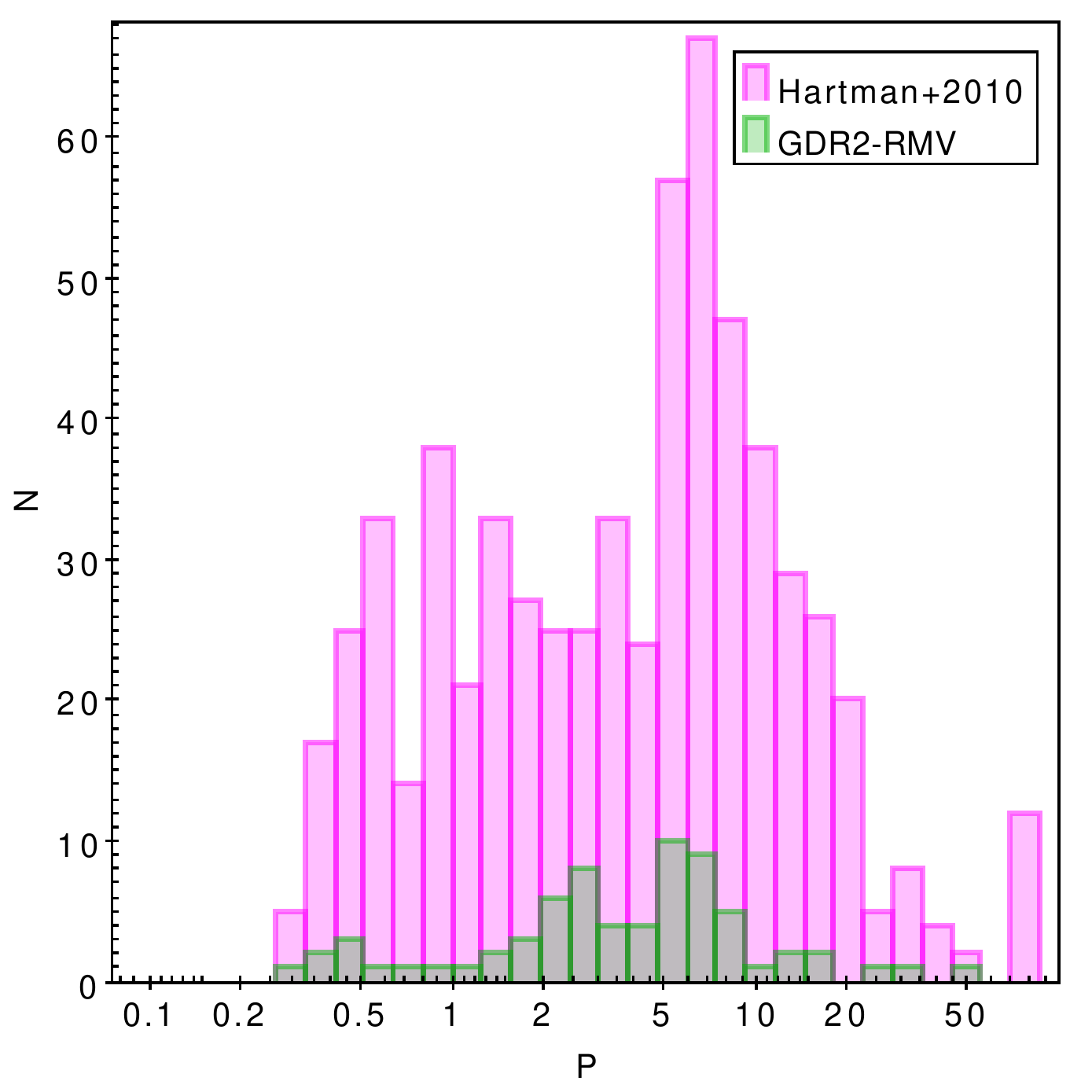}
\includegraphics[width=0.44\textwidth]{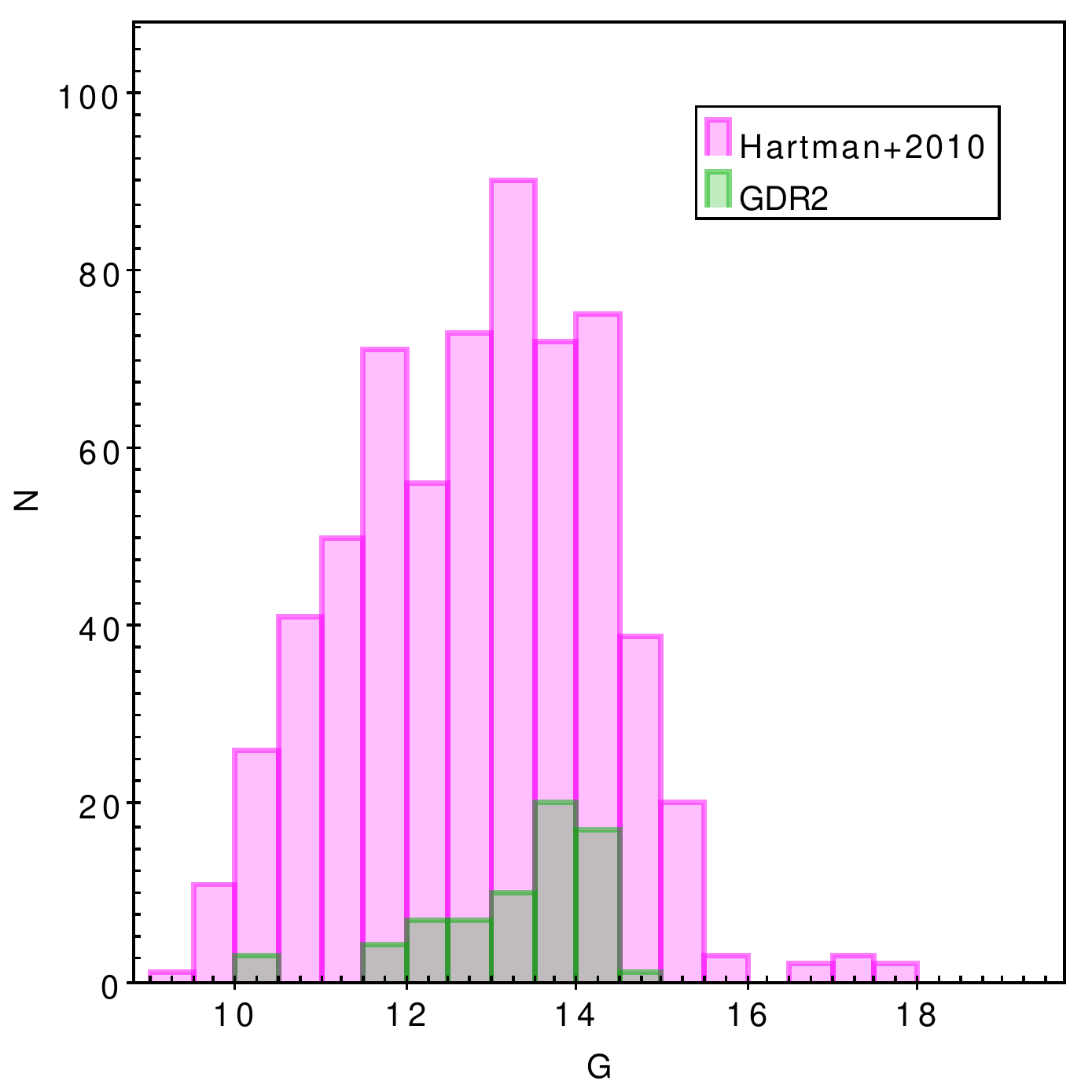}
\includegraphics[width=0.44\textwidth]{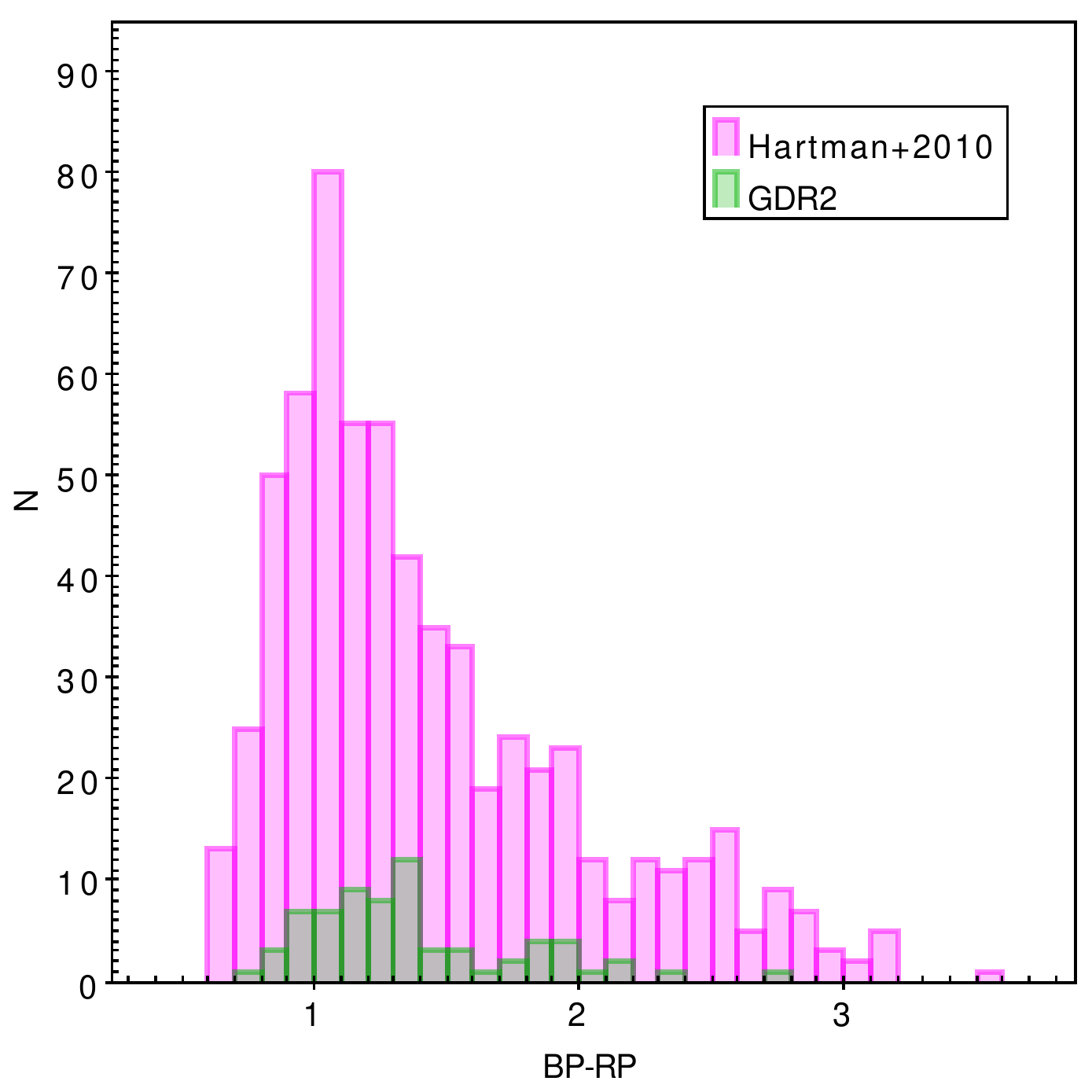}
\caption{
Comparison of the $P$, $G$, and $BP-RP$ distributions between the \gaia\ DR2 BY\,Dra and the \cite{2010MNRAS.408..475H} samples in the overlapping \texttt{HEALPix} and in common $P$, $G,$ and $BP-RP$ ranges.
$P$ is given in day units.}
\label{fig:Hartman_common_comparison}
\end{center}
\end{figure*}

Then we take the $P$, $G,$ and \bmr\ distributions in the overlapping \texttt{HEALPix} of the two samples into account.
The \gaia\ sample contains fainter sources than the Hartman sample.
On the other hand, the \gaia\ sample has a more restrictive lower cut of 0.3 d in $P$ and of 0.6 at the bluer end in \bmr.
By restricting the comparison samples to similar $P$, $G$, and \bmr\ ranges, we obtain the comparison map shown in Fig.\,\ref{fig:Hartman_common_comparison}.
The count ratio, in the remaining \texttt{HEALPix} with stars in common, ranges from 0.02 to 2, with a median of 0.14, in 36 common level 6 \texttt{HEALPix}.
From this we roughly estimate that the \gaia\ completeness upper limit is 14\% in the common \texttt{HEALPix}.
Furthermore, if we assume that this completeness percentage is representative for the whole sky, considering that the \gaia\ sample covers 18593 out of 49152 level-6 \texttt{HEALPix}, we estimate an upper limit of 5\% completeness for the whole sky down to $G\approx16.5$.

At the other extreme, for what matters here, we assume that all late-type dwarfs observed by Gaia have surface inhomogeneities and therefore should display rotational modulation, but Gaia does not reveal it.
We then compare the \cite{2010MNRAS.408..475H} sample with all late-type dwarfs observed by Gaia in the same FoV (same level 6 \texttt{HEALPix}).
In this way, we find a median count ratio of 0.14, from which we estimate a lower limit of 14\% for the completeness of the \cite{2010MNRAS.408..475H} sample and therefore a lower limit of 0.7\% for the \gaia\ DR2 BY\,Dra sample over the whole sky.

The expected main sources of uncertainty in this simplified completeness estimate are those related to the dependence of the detection efficiency (Eq.\,\ref{eq:efficiency_def}) on the ecliptic coordinate and on $P$.
Figs.\,\ref{fig:map} and \ref{fig:Pdistr} give an idea of the problem: the distribution of detected $P$ depends on the ecliptic coordinates, and the preliminary data contained in \gaia\ DR2 are still insufficient for a meaningful empirical determination of the $Q$ dependencies.
Regarding the dependence on $P$, taking an existing survey as reference would require, in principle, that
such a survey has at least a known $P$ detection efficiency.
This is not our case, but taking a survey that explored just a sky region containing the Pleiades cluster ($\approx$\,120\,Myr old) is expected to have no great consequences on our overall percentage completeness estimate.
Considering both cluster members and non-members, as available in the \cite{2010MNRAS.408..475H} catalogue, provides a sample  sufficiently rich to make our rough estimate meaningful even without a detailed knowledge of the detection efficiency of the
reference catalogue $P.$    
Further work is required in this direction before the final \gaia\ data release.

\section{Catalogue retrieval}

Here we present some examples of ADQL queries to be used in the web interface to the \gaia\  DR2 archive (\href{https://gea.esac.esa.int/ archive/}{https://gea.esac.esa.int/ archive/}) to retrieve the BY\,Dra data.

Get the first 1000 stars in the rotational modulation table: 
\begin{verbatim}
SELECT TOP 1000 * 
FROM gaiadr2.vari_rotation_modulation
\end{verbatim}

Get the position, parallax, and link to the light curves for the first 1000 sources in the rotational modulation table:
\begin{verbatim}
SELECT TOP 1000 gs.source_id, 
gs.ra, 
gs.dec, 
gs.parallax, 
epoch_photometry_url 
FROM gaiadr2.gaia_source as gs 
INNER JOIN gaiadr2.vari_rotation_modulation as rotmod 
ON gs.source_id=rotmod.source_id
\end{verbatim}

Get the time-series statistics for the first 1000 sources in the rotational modulation table:
\begin{verbatim}
SELECT top 1000 tsstat.* 
FROM gaiadr2.vari_time_series_statistics as tsstat 
INNER JOIN gaiadr2.vari_rotation_modulation as rotmod 
ON tsstat.source_id=rotmod.source_id
\end{verbatim}

Get the final rotation period and amplitude of the first 100 sources in the rotational modulation table:
\begin{verbatim}
SELECT top 100 source_id, 
best_rotation_period, 
max_activity_index 
FROM gaiadr2.vari_rotation_modulation
\end{verbatim}

Select parallax, magnitudes, period, and modulation amplitude for the first 100 sources in the rotational modulation table:
\begin{verbatim}
SELECT TOP 100 gs.source_id, 
gs.parallax, 
gs.phot_g_mean_mag, 
gs.phot_bp_mean_mag, 
gs.phot_rp_mean_mag, 
rm.best_rotation_period, 
rm.max_activity_index 
FROM gaiadr2.gaia_source AS gs 
INNER JOIN gaiadr2.vari_rotation_modulation AS rm 
ON gs.source_id=rm.source_id
\end{verbatim}

\end{document}